\documentclass[a4paper,11pt]{article}
\usepackage{jheppub, siunitx, hyperref, booktabs, subcaption, bm, xcolor, adjustbox, multirow}                    
\usepackage[capitalise]{cleveref}  
\usepackage[T1]{fontenc} 
\usepackage[acronym]{glossaries}

\DeclareSIUnit{\nothing}{\relax} 
\DeclareSIUnit\barn{b}
 
\DeclareUnicodeCharacter{2212}{-}

\newcommand{\neutralino}[1]{\Tilde{\chi}_{#1}^0}

\newcommand{\squark}[1]{\widetilde{{#1}}}
\newcommand{\antisquark}[1]{\squark{#1}^*}
\newcommand{\software}[1]{\texttt{#1}}
\newcommand{\llh}{\mathcal{L}} 

\newacronym{PDF}{PDFs}{parton distribution functions}
\newacronym{LO}{LO}{leading-order}
\newacronym{NLO}{NLO}{next-to-leading-order}
\newacronym{NLL}{NLL}{next-to-leading-logarithmic}
\newacronym{NNLO}{NNLO}{next-to-next-to-leading-order}
\newacronym{NNLL}{NNLL}{next-to-next-to-leading-logarithmic}
\newacronym{aNNLO}{aNNLO}{approximate next-to-next-to-leading-order}

\newacronym{SUSY}{SUSY}{supersymmetric}
\newacronym{SM}{SM}{Standard Model}
\newacronym{BSM}{BSM}{Beyond the Standard Model}
\newacronym{QCD}{QCD}{Quantum Chromo Dynamics}
\newacronym{MSSM}{MSSM}{Minimal Supersymmetric Standard Model}
\newacronym{WIMP}{WIMP}{weakly interacting massive particle}

\newacronym{PAD}{PAD}{Public Analysis Database}
\newacronym{LSP}{LSP}{lightest supersymmetric particle}
\newacronym{SR}{SR}{signal region}
\newacronym{CL}{CL}{confidence level of exclusion}
\newacronym{llr}{LLR}{logarithmic likelihood ratio}
\newacronym{MA5}{\software{MA5}}{\software{MadAnalysis5  1.10.2}}
\newacronym{DJR}{DJR}{differential jet rate}
\newacronym{ME}{ME}{matrix element}

\title{Combination and Reinterpretation of LHC SUSY Searches}

\author[a]{Alexander Feike}
\author[b,c]{\!\!, Juri Fiaschi}
\author[d]{\!\!, Benjamin Fuks}
\author[a]{\!\!, Michael Klasen}
\author[a]{and Alexander~Puck~Neuwirth}

\affiliation[a]{Institut  für  Theoretische  Physik,  Universität  Münster,  Wilhelm-Klemm-Straße 9, 48149 Münster, Germany}
\affiliation[b]{Universit\`a degli Studi di Milano-Bicocca, Department of Physics “Giuseppe Occhialini”, \& INFN, Sezione di Milano-Bicocca, Piazza della Scienza 3, Milano 20126, Italy}
\affiliation[c]{Department of Mathematical Sciences, University of Liverpool, Liverpool L69 3BX, United Kingdom}
\affiliation[d]{Laboratoire de Physique Th\'eorique et Hautes \'Energies (LPTHE), UMR 7589, Sorbonne Universit\'e et CNRS, 4 place Jussieu, 75252 Paris Cedex 05, France}

\emailAdd{alex.feike@uni-muenster.de}
\emailAdd{juri.fiaschi@unimib.it}
\emailAdd{fuks@lpthe.jussieu.fr}
\emailAdd{michael.klasen@uni-muenster.de}
\emailAdd{alexander.neuwirth@uni-muenster.de}

\abstract{
To maximise the information obtained from various independent new physics searches conducted at the LHC, it is imperative to consider the combination of multiple analyses. To showcase the exclusion power gained by combining signal regions from different searches, we consider a simplified scenario inspired by supersymmetry, with all particles but one squark flavour and a bino-like neutralino decoupled. The corresponding signal therefore comprises strong squark pair production, associated squark-neutralino production, as well as weak neutralino pair production. We find that considering the associated and strong production mechanisms together significantly impacts mass limits, while contributions from the weak production are insignificant in the context of current exclusion limits. In addition, we demonstrate that the combination of uncorrelated signal regions as assessed from the recent \software{TACO} approach substantially pushes exclusion limits towards higher masses, relative to the bounds derived from the most sensitive individual analyses.
}

\preprint{MS-TP-23-49}

\begin{document} 

\maketitle
\flushbottom

\section{Introduction}
\label{sec:intro}
Observations like the rotation curves of remote stars or gravitational lensing strongly indicate that our Universe includes more than just visible matter and the neutrinos of the \gls*{SM}. A promising idea to explain this surplus of mass is to introduce a stable, or at least long living, \gls*{WIMP} in a  \gls*{BSM} theory. One appealing possibility for such a construction is to extend the theory symmetry group with fermionic operators. This leads to the concept of \gls*{SUSY} quantum field theories, which additionally maximally extend the Poincar\'{e} group~\cite{PhysRev.159.1251, Haag:1974qh, Wess:1974tw}. To comply with dark matter considerations, most \gls*{SUSY} models, and in particular the minimal supersymmetric realisation called the \gls*{MSSM}, further assume the conservation of a multiplicative quantum number dubbed $R$-parity~\cite{Farrar:1978xj}. This results in phenomenologically viable SUSY spectra where the \gls*{LSP} is electrically neutral~\cite{Jungman:1995df}.

Consequently, intensive searches for \gls*{SUSY} particles have been taken up at high-energy particle colliders like the Large Hadron Collider (LHC) at CERN. The ATLAS and CMS experiments at the LHC initially focused on signatures of the strong production of squarks and gluinos~\cite{ATLAS:2011xeq,CMS:2011xek}, and next extended their searches to incorporate weak production channels, \textit{i.e.}\ the pair production of neutralinos, charginos and sleptons~\cite{ATLAS:2012uah, ATLAS:2014ikz, CMS:2012hnc, CMS:2013bda}. Although no evidence for \gls*{SUSY} has been found yet, it remains a cutting-edge research topic until today. Recent experimental analyses at the LHC~\cite{ATLAS:2021fbt, ATLAS:2023afl, ATLAS:2024fyl, CMS:2023xlp, CMS:2023yzg} pushed the lower bounds on the SUSY masses deep into the \SI{}{\tera \electronvolt} regime, the exact limit depending on the scenario considered. However, in certain model scenarios such as when the \gls*{SUSY} spectrum is compressed, these limits may not apply, allowing for potential escape routes from the constraints imposed experimentally. While limits are usually cast in the context of simplified models inspired by the \gls*{MSSM} and in which only a few \gls*{SUSY} states are light, data can be reinterpreted in different, non-minimal and more realistic, scenarios. In this case, typical \gls*{SUSY} signals often comprise several components. For instance, in the framework of a scenario featuring light squarks and a neutralino \gls*{LSP}, a \gls*{SUSY} signature made of jets and missing transverse energy could arise from both squark-pair and associated squark-neutralino production, in contrast with simplified model setups in which only a single production channel is considered. The mutual impact of these two processes on \gls*{SUSY} exclusions computed from the signal region of a specific LHC analysis providing the best expectation has been recently examined~\cite{Lara:2022new}. Light was shed on the existence of a non-trivial improvement on the \gls*{SUSY} bounds, which finds its origin in a more accurate modelling of the \textit{full} \gls*{SUSY} signal by incorporating all its potentially relevant components. 

To further profit from the already collected data and from all searches for multiple jets and missing energy, it is necessary to also add pure weak production channels in which a pair of electroweakinos is produced. Such weak processes could indeed be relevant in specific regions of the \gls*{SUSY} parameter space. Furthermore, the new physics search programme at the LHC includes a variety of searches for jets and missing transverse energy, each leveraging slightly different handles on the signal. It is therefore crucial to combine \glspl*{SR} of any specific analysis as well as different analyses, although this must be achieved in a statistically sound approach without double counting any effect. Experimental collaborations have recently made efforts to provide information on the statistical models used for limit setting, including correlations between \glspl*{SR}, either exactly~\cite{ATL-PHYS-PUB-2019-029} or approximately~\cite{CMS-NOTE-2017-001}. However, comprehensive information is only available for a limited number of LHC analyses. On the other hand, the code \software{TACO}~\cite{Araz:2022vtr} tackles the problem of combining \glspl*{SR} from different channels in an approximate manner, by determining the existing correlation between \glspl*{SR} of possibly different LHC analyses for a specific set of signal events. This information then provides a way to compute exclusion limits from the best possible and meaningful combination of a given set of \glspl*{SR}, that is in practice obtained by means of a graph-based algorithm. While this approach offers a first insight into the gain in sensitivity achieved by combining \glspl*{SR} from various analyses, it ignores correlated systematic uncertainties like those originating from parton densities, luminosity measurements or limitations arising from overlapping signal regions of a particular analysis and control regions from other analyses.

In this work, we calculate the best estimation to date of limits on non-minimal \gls*{SUSY} models with squarks and electroweakinos, complementing earlier work assessing the gain in sensitivity obtained from the combination of various ATLAS and CMS searches for electroweakinos~\cite{Altakach:2023tsd} and third-generation squarks~\cite{Alguero:2022gwm}. These bounds are obtained not only from simulations of the full corresponding \gls*{SUSY} signal, but also from the combination of several searches for \gls*{SUSY} relying on jets and missing transverse energy. To this aim, we consider a simplified \gls*{SUSY} scenario in which the set of relevant superpartners is restricted to one neutralino and one squark, as described in \cref{sec:theory}. In this section, we also discuss the different processes contributing to the signal. Limits are determined from the toolchain setup introduced in \cref{sec:analysis}, that allows for event generation, detector simulation, cross section calculation, LHC recasting and \gls*{SR} combination. The results displayed in \cref{sec:results} then showcase the resulting gain in exclusion power. Finally, we conclude in \cref{sec:concl}. 

\section{Theoretical setup and hard-scattering signal simulation}
\label{sec:theory}
We consider a simplified model inspired by the \gls*{MSSM}, in which the \gls*{SM} is extended by one squark flavour $\squark{q} \equiv \squark{u}_R$ and one neutralino state $\neutralino{1}$, all other superpartners having masses of \SI{30}{\tera \electronvolt} and being thus decoupled. Furthermore, the neutralino mixing matrix is taken to be diagonal, so that the lightest $\neutralino{1}$ state is effectively bino-like. In this new physics parametrisation, the mass of the squark $m_{\squark{q}}$ and that of the neutralino $m_{\neutralino{1}}$ are free parameters, and their values are imposed to satisfy the condition $m_{\squark{q}} > m_{\neutralino{1}}$. This ensures that the $\neutralino{1}$ state is the \gls*{LSP}, and therefore a good candidate for dark matter (thanks to $R$-parity conservation). The associated collider signal therefore includes three processes: the strong production of a pair of squarks ($p\,p \rightarrow \squark{q}\, \antisquark{q}$), the associated production of a neutralino and a squark ($p\,p \rightarrow \squark{q}\, \neutralino{1}  + \mathrm{H.c.}$), and the weak production of a pair of neutralinos ($p\,p \rightarrow \neutralino{1}\, \neutralino{1}$). All superpartners different from the $\squark{q}$ and $\neutralino{1}$ states being decoupled, the squark $\squark{q}$ is unstable and always decays promptly via the process $\squark{q} \rightarrow \neutralino{1} q$. The global signature of the signal, once squark decays are accounted for, is therefore made of multiple jets and missing transverse energy carried away by the stable neutralino states. 

In this way, signal simulation is achieved by means of \software{MadGraph5\_aMC@NLO 3.5.1}~\cite{Alwall:2014hca}. Making use of the UFO~\cite{Darme:2023jdn} implementation of the \gls*{MSSM}~\cite{Duhr:2011se} obtained with \software{FeynRules}~\cite{Christensen:2009jx, Alloul:2013bka}, we independently generate hard-scattering events for the three considered sub-processes. To this aim, we convolute \gls*{LO} matrix elements featuring up to two additional jets (from either initial-state or final-state radiation) with the  \software{MSHT20lo\_as130}~\cite{Bailey:2020ooq} \gls*{LO} set of \gls*{PDF} provided by \software{LHAPDF 6.5.4}~\cite{Buckley:2014ana}. To maintain consistency in the analysis, it is imperative to avoid double counting any contribution across the three production modes. Thus, we manually prohibit intermediate squarks from being on-shell in any $2\to3$ or $2\to4$ diagram (\textit{i.e.}\ with one or two QCD emissions). The produced events are next re-weighted according to $K$-factors defined by the ratio of total rates including state-of-the-art higher-order corrections obtained with \software{Resummino 3.1.2}~\cite{Fiaschi:2023tkq} and \software{NNLL-fast 2.0}~\cite{Beenakker:2016lwe} to \gls*{LO} predictions computed with these two codes. 

Neutralino pair production cross sections are calculated with \software{Resummino}.\footnote{Executed via the \software{Python} interface \software{HEPi}~\cite{HEPi}.} The code combines, following the standard threshold resummation formalism~\cite{Sterman:1986aj, Catani:1989ne, Vogt:2000ci}, matrix elements including \gls*{aNNLO} corrections in QCD~\cite{Beenakker:1999xh, Fiaschi:2020udf} with the resummation of soft-gluon radiation at the \gls*{NNLL} accuracy~\cite{Debove:2010kf, Fuks:2012qx, Fiaschi:2018hgm, Fiaschi:2020udf}. We additionally use  \software{Resummino} to compute associated neutralino-squark production rates by matching \gls*{NLO} matrix elements~\cite{Baglio:2021zjm} with threshold resummation at the \gls*{NLL} accuracy~\cite{Fiaschi:2022odp}. Finally, total cross sections for squark pair production are calculated with \software{NNLL-fast}. The latter combines \gls*{aNNLO} matrix elements~\cite{Beenakker:1996ch, Langenfeld:2009eg}, soft-gluon resummation at \gls*{NNLL} in the absolute threshold limit, and the resummation of Coulomb gluons with an NLO Coulomb potential and bound-state contributions~\cite{Kulesza:2008jb, Kulesza:2009kq, Beenakker:2009ha, Beenakker:2011sf, Beenakker:2013mva, Beenakker:2014sma}.\footnote{While \software{NNLL-fast} assumes a tenfold-degenerate squark spectrum, the dependence of the cross section on the nature of the squarks factorises when the gluino is decoupled. Moreover, \software{NNLL-fast} is unable to calculate rates at precision below \gls*{NLO}. We therefore utilise its predecessor \software{NLL-fast}~\cite{Beenakker:2015rna} and its predictions at \gls*{NLO} and \gls*{LO} to calculate a global $K$ factor defined by $K = (\sigma_\mathrm{NLO}/\sigma_\mathrm{LO})_\texttt{NLL-fast} \times (\sigma_\mathrm{aNNLO+NNLL}/\sigma_\mathrm{NLO})_\texttt{NNLL-fast}$.}

\begin{figure}
    \centering
    \includegraphics[width=0.75\textwidth]{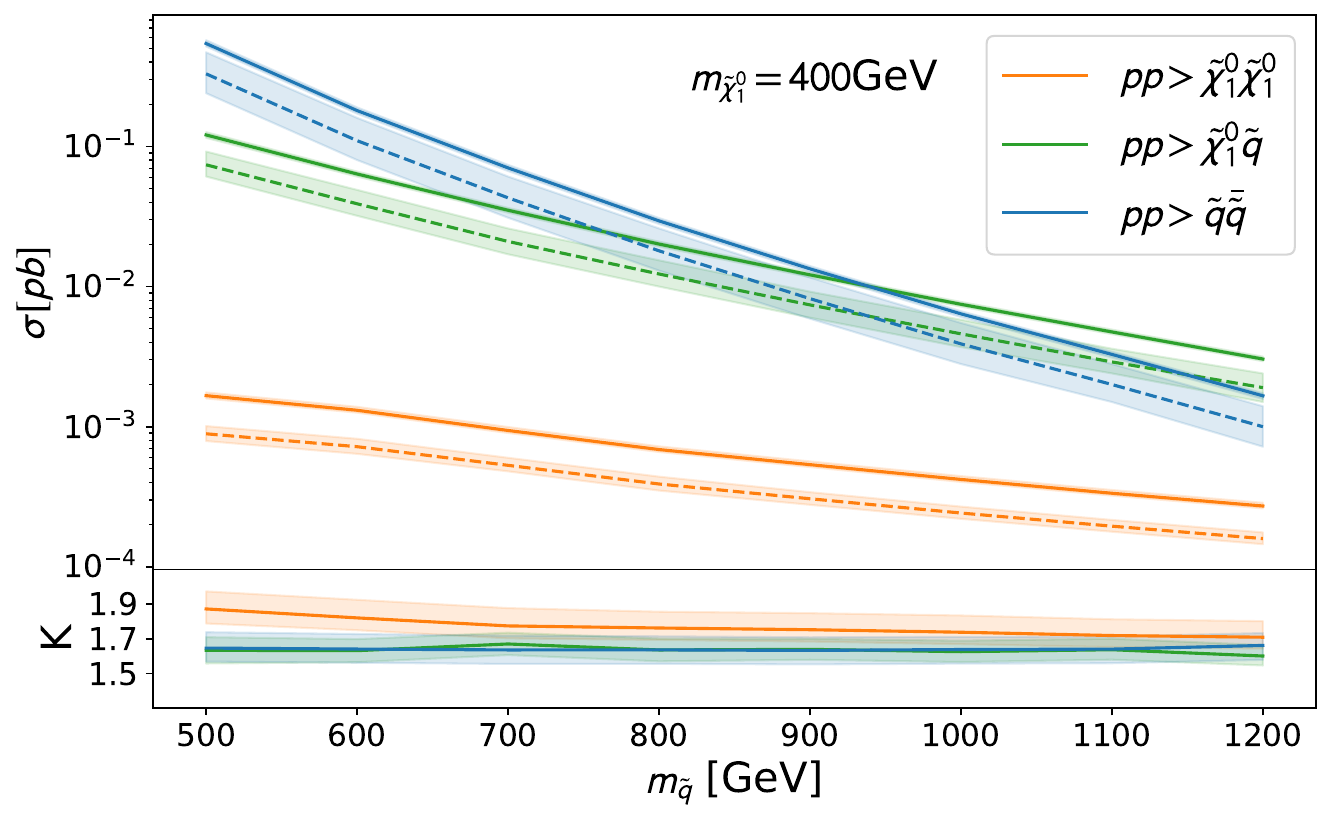}
    \caption{Leading-order (dashed) and higher-order resummed (solid) cross section predictions in $pp$ collisions at the LHC with a centre-of-mass energy of 13 TeV for the three considered processes (upper panel) and associated $K$-factors (lower panel) including scale uncertainties, for a scenario featuring a fixed neutralino mass of $m_{\neutralino{1}}=\SI{400}{\giga \electronvolt}$ and a varying squark mass $m_{\squark{q}}$.}
    \label{fig:xsec}
\end{figure}
As an illustration of the impact of the $K$-factors for the three processes considered, we focus on a scenario in which the mass of the lightest neutralino is fixed to $m_{\neutralino{1}}=\SI{400}{\giga \electronvolt}$, and we then show in \cref{fig:xsec} the dependence of the three production cross sections on the squark mass $m_{\squark{q}}$. We present, in the upper panel of the figure, predictions at \gls*{LO} (dashed lines) and after including state-of-the-art higher-order corrections in QCD (solid lines).
As expected, the relative uncertainties stemming from scale variation, that are also shown on the figure, decrease significantly when adding resummation contributions to the LO cross sections. We remind that a thorough error handling including PDF uncertainties for resummed cross sections can be found in \cite{Fiaschi:2022odp,Fiaschi:2023tkq} for the weak and associated production channels. In the following analysis, the experimental errors nevertheless dominate such that the theoretical uncertainties are only shown to present an overview.
In the lower inset of the figure, we additionally display the ratio between the two. For light squarks, the cross section corresponding to strong squark pair production ($\sigma \propto \alpha_S^2$) is the highest, while the semi-weak/semi-strong associated neutralino-squark production cross section ($\sigma \propto \alpha_S \alpha$) and the weak neutralino pair production cross section ($\sigma \propto \alpha^2$) are suppressed by about one and three orders of magnitude, respectively. As the squark gets heavier, the strong cross section decreases quickly due to phase space suppression, and is eventually overcome by the semi-weak process rates. The turnover occurs, in our example with $m_{\neutralino{1}}=\SI{400}{\giga \electronvolt}$, at around $m_{\squark{q}}\approx \SI{920}{\giga \electronvolt}$. By virtue of the same effect, at even higher squark masses the purely weak process become increasingly relevant. $K$-factors are generally large and exhibit little dependence on the squark mass. Such large $K$-factors (especially for small squark masses in the weak $p\,p \to \neutralino{1}\, \neutralino{1}$ channel) have been cross-checked with \software{Prospino}~\cite{Beenakker:1996ed}, and are also outlined in~\cite{Gao:2002is}.

Taking into account current exclusions that slowly reach the \SI{}{\tera \electronvolt} regime, all three processes could have comparable cross sections in any given realistic scenario. Their contribution in the modelling of the \gls*{SUSY} signal should therefore be correctly incorporated, which further motivates the necessity of their unified analysis as achieved in this study.

\section{Analysis setup}
\label{sec:analysis}

\begin{figure}
    \centering
    \includegraphics[width=0.8\textwidth]{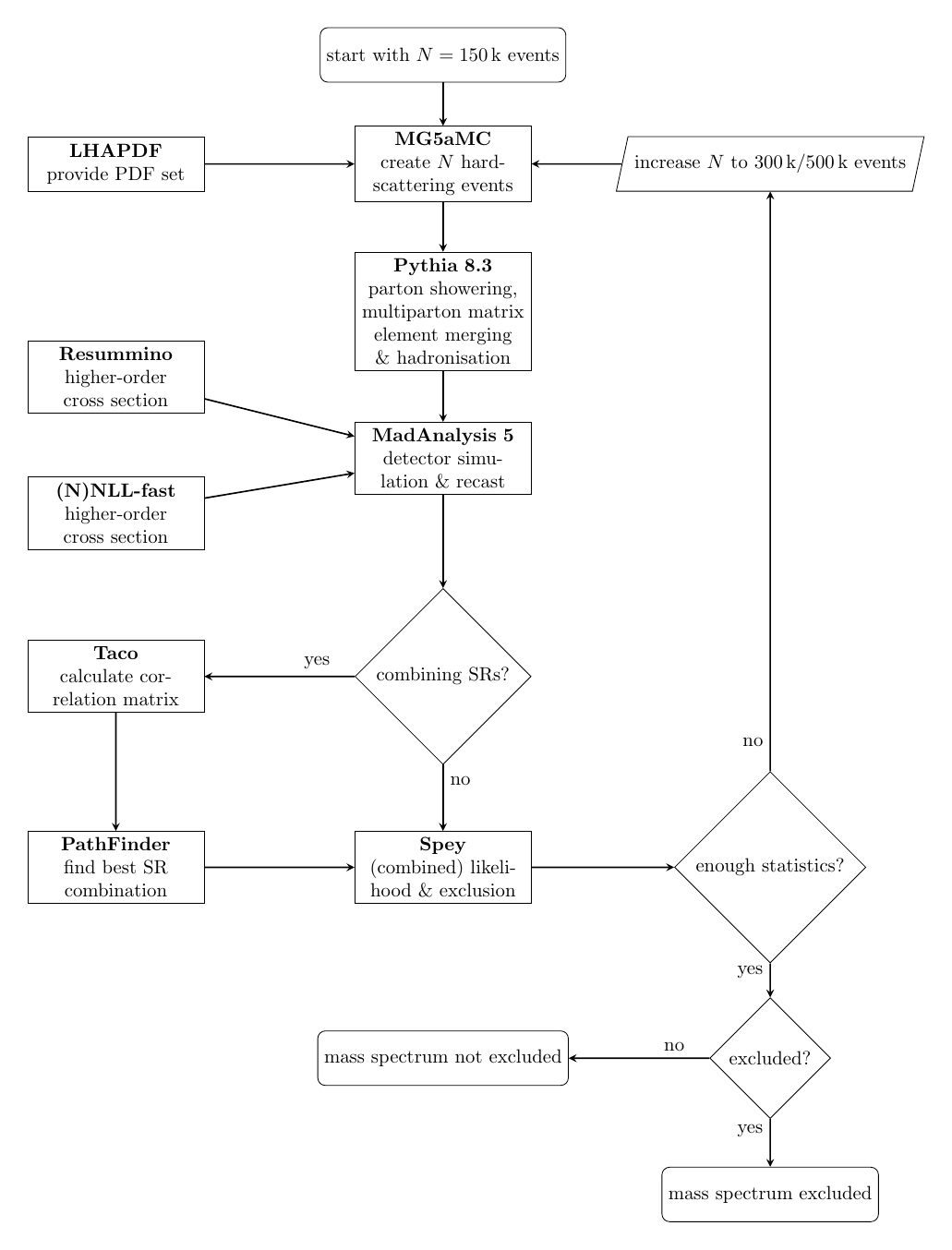}
    \caption{Flowchart diagram of the used toolchain.}
    \label{fig:flowchart}
\end{figure}
The computation of the exclusion associated with each point in the model parameter space follows the toolchain outlined in \cref{fig:flowchart}. For every combination of squark and neutralino masses, we generate \SI{150}{\kilo \nothing} events with \software{MadGraph5\_aMC@NLO 3.5.1} (\software{MG5aMC}) which is then interfaced with \software{Pythia 8.306}~\cite{Bierlich:2022pfr} for parton showering and hadronisation. Furthermore, we utilise \software{Pythia} to merge event samples featuring a different final-state partonic multiplicity following the \software{CKKW-L} algorithm~\cite{Catani:2001cc, Lonnblad:2001iq, Lonnblad:2011xx}, with a merging scale set to one quarter of the \gls*{SUSY} hard scale defined by the average mass of the produced heavy particles. This scale is thus equal, for the three considered sub-processes, to
\begin{equation}
    Q_{\text{MS}} = 
    \begin{cases}
       \frac{1}{4} m_{\neutralino{1}} & \text{for }pp\rightarrow \neutralino{1} \neutralino{1}\\
       \frac{1}{8} (m_{\squark{q}} + m_{\neutralino{1}}) & \text{for }pp\rightarrow \neutralino{1}\squark{q} + \neutralino{1}\antisquark{q} \\
       \frac{1}{4} m_{\squark{q}} & \text{for }pp\rightarrow \squark{q} \antisquark{q}
    \end{cases}
    \, . \label{eq. merging scale}
\end{equation}
We refer to \cref{appx: event generation} for additional details on the merging procedure. The event weights are then re-scaled so that total rates encompass higher-order resummed corrections, as described in ~\cref{sec:theory}.

We simulate the response of the LHC detectors using \gls*{MA5}~\cite{Conte:2012fm, Conte:2014zja, Conte:2018vmg}, which relies either on \software{Delphes 3}~\cite{deFavereau:2013fsa} or its built-in \software{SFS} simulator~\cite{Araz:2020lnp}. For each LHC analysis considered (detailed in \cref{sec:recast}), we employ a tuned detector parametrisation and the validated analysis implementation within \gls*{MA5}. This allows us to assess the selection efficiency for our signal across all \glspl*{SR} of all analyses examined, and to construct an acceptance matrix where each event corresponds to a row and each \gls*{SR} to a column. We next utilise the software \software{TACO}~\cite{Araz:2022vtr} to convert the acceptance matrix into a correlation matrix. This conversion enables the determination of which combination of uncorrelated \glspl*{SR} yields the highest exclusion power. To achieve this, we employ a generalised version of the \software{PathFinder} software, initially implemented within \software{TACO}.\footnote{This generalised version of the \software{PathFinder} software can be found at \url{https://github.com/J-Yellen/PathFinder}.} The obtained information is next passed to the statistical tool \software{SPEY}~\cite{Araz:2023bwx}, which uses a variety of likelihood-based prescriptions for hypothesis testing and the derivation of the \gls*{CL} of a single SR, or of a combination of multiple \glspl*{SR}. Since the number of events surviving the cuts varies from \gls*{SR} to \gls*{SR}, if the number of events populating the \gls*{SR} with the highest exclusion power is too low ($\lesssim50$ events), we regenerate events with increased statistics, first considering \SI{300}{\kilo \nothing} events, and next \SI{500}{\kilo \nothing} events if \SI{300}{\kilo \nothing} events are still insufficient. 

Iterating this procedure scanning over the relevant mass points in the model parameter space, we obtain the \SI{95}{\percent} \gls*{CL} exclusion contour for the individual processes, as well as for their combination. The exclusion limits are calculated at first for the individual processes within each analysis, selecting the most sensitive \gls*{SR} as the one with the highest obtained exclusion value. This is further generalised to obtain the combined exclusion limits for the three sub-processes together. Further practical details (including information on the analyses considered and our limit setting procedure) are collected in \cref{sec:recast} and \cref{sec:combination}.

\subsection{Recast of single experimental analyses}
\label{sec:recast}

All events generated, alongside their associated (resummed) cross sections, are passed to \gls*{MA5} to reinterpret the results of four Run~2 ATLAS and CMS analyses at a centre-of-mass energy of $\sqrt{s}=\SI{13}{\tera \electronvolt}$. These analyses respectively account for integrated luminosities of $\SI{139}{\femto\barn^{-1}}$ and $\SI{137}{\femto\barn^{-1}}$ of data. Specifically, we investigate four searches that target jets and missing transverse momentum and that include \glspl*{SR} focusing not only on a monojet signature but also on a topology allowing for multiple hard jets. These searches are identified as ATLAS-EXOT-2018-06~\cite{ATLAS:2021kxv}, ATLAS-CONF-2019-040~\cite{ATLAS:2020syg}, CMS-SUS-19-006~\cite{CMS:2019zmd}, and CMS-EXO-20-004~\cite{CMS:2021far}, and they have demonstrated the highest sensitivity to the considered simplified scenario. Details regarding their integration into \gls*{MA5}, along with corresponding validation notes, can be found on the \gls*{MA5} \software{dataverse}~\cite{DVN/NW3NPG_2021, DVN/REPAMM_2023, DVN/4DEJQM_2020, DVN/IRF7ZL_2021}, on the \gls*{MA5} \gls*{PAD}~\cite{Dumont:2014tja}, as well as in the works~\cite{Araz:2019otb, Kim:2020nrg, Mrowietz:2020ztq, CMS:2021far, Fuks:2021zbm}. The preselection criteria defining these four analyses are briefly summarised in \cref{tab:analyses cuts}.

\begin{table}
    \centering\renewcommand{\arraystretch}{1.3}\setlength{\tabcolsep}{12pt}
     \begin{adjustbox}{max width=.97\textwidth}
     \begin{tabular}{l c c c c}
        \multirow{2}{*}{Cuts} & ATLAS & ATLAS & CMS & CMS\\
         & EXOT-2018-06 & CONF-2019-040 & SUS-19-006 & EXO-20-004 \\
         \midrule
         veto &  $e$, $\mu$, $\tau$, $\gamma$ & $e$, $\mu$ & $e$, $\mu$, $\gamma$& $e$, $\mu$, $\tau$, $\gamma$, $b$-jet  \\
         $N_j$ & $\geq 1$& $\geq 2$ & $\geq 2$ &  $\geq 1$\\
         $E_T^{\text{miss}}$ & $>\SI{200}{\giga \electronvolt}$ & $>\SI{300}{\giga \electronvolt}$ & - & $>\SI{250}{\giga \electronvolt}$ \\
         $|\eta|$ & $<2.4$ & - & $<2.4$ & $<2.4$ \\
         $p_T(j_1)$ & $>\SI{150}{\giga \electronvolt}$ & $>\SI{200}{\giga \electronvolt}$ & - & $>\SI{100}{\giga \electronvolt}$  \\
         $p_T(j_2,...,j_{N_j})$ & $>\SI{30}{\giga \electronvolt}$ & $>\SI{50}{\giga \electronvolt}$ & $>\SI{30}{\giga \electronvolt}$  & $>\SI{20}{\giga \electronvolt}$  \\
         $\Delta \Phi(\text{jet}, \mathbf{p}_T^{\text{miss}})$ & >0.4  & $> 0.2$& >0.5 & >0.5 \\
         $m_{\text{eff}}$ & - & $>\SI{800}{\giga \electronvolt}$ & - & - \\
         $H_T$ & - & - & >\SI{300}{\giga \electronvolt} & - \\
         $|\vec{H}_T^{\text{miss}}|$ & - & - & >\SI{300}{\giga \electronvolt} & - \\
    \end{tabular}
    \end{adjustbox}
    \caption{Summary of the event selection cuts of the four analyses ATLAS-EXOT-2018-06, ATLAS-CONF-2019-040, CMS-SUS-19-006 and CMS-EXO-20-004. Requirements include vetos on specific objects, as well as selections on the number of jets $N_j$, the amount of missing transverse energy $E_T^{\text{miss}}$, the pseudo-rapidity $|\eta|$ and transverse momentum $p_T$ of the different jet candidates, the angular separation between the jets and the missing transverse momentum $\Delta \Phi(\text{jet}, \mathbf{p}_T^{\text{miss}})$, and the global observables defined in Eq.~\eqref{eq:globalobs}.}
    \label{tab:analyses cuts}
\end{table}

These cuts have been devised to possibly observe the targeted \gls*{BSM} signal, which is expected to be characterised by a significant production of events containing energetic jets, no leptons and a substantial amount of missing transverse energy $E_T^{\text{miss}}$. Although the acceptance cuts implemented in the four analyses are largely similar, they define different signal regions, thereby yielding varying sensitivities across the model parameter space. In particular, we firstly notice that the requirements on the minimal number of jets and their minimal transverse momentum $p_T(j)$ vary slightly. In this way, the CMS-EXO-20-004 analysis is sensitive to a softer monojet-like signature (thanks to lower cut thresholds) while the ATLAS-CONF-2019-040 search is blind in this region. Secondly, global cuts on the effective mass $m_\mathrm{eff}$ and the visible ($H_T$) and invisible ($\vec{H}_T^{\text{miss}}$) hadronic activity are only imposed in some of the considered analyses, that define these variables as
\begin{equation}\label{eq:globalobs}\begin{split}
    m_{\text{eff}}=E_T^{\text{miss}} + \sum \limits_{p_T>\SI{50}{\giga \electronvolt}}p_T(j)\,,\qquad
    H_T = \sum \limits_{|\eta|<2.4}p_T(j)\, ,\qquad
    \vec{H}_T^{\text{miss}} = \sum \limits_{|\eta|<5}\vec{p}_T(j)\, .
\end{split}\end{equation}
The cuts on $H_T$ and $\vec{H}_T^{\text{miss}}$ implemented in the CMS-SUS-19-006 analysis imply a focus on a signal topology in which the hadronic activity originates from a large number of not necessarily very hard jets. In contrast, the ATLAS-EXOT-2018-06 search requires only one highly energetic jet, and optionally additional softer jets.

Using \software{MA5}, we calculate the number of signal events $n_s$ expected to populate each \gls*{SR} of the four analyses. This information, together with the corresponding number of experimentally observed events $n_{\text{obs}}$, the number of expected \gls*{SM} background events $n_{\text{b}}$ and the associated uncertainty $\Delta n_{\text{b}}$ is passed to \software{SPEY} for statistical analysis. Using a general composite likelihood approach which employs Poisson distributed data, Gaussian-distributed nuisance and the so-called test statistic $\tilde{q}_\mu$~\cite{Cowan:2010js}, \software{SPEY} computes separate \glspl*{CL} exclusion for each \gls*{SR} of each analysis and for the three sub-processes. To get exclusion limits for the combination of the three processes, we consider that each \gls*{SR} is populated by a number of events given by
\begin{equation}
   n_s=n_s^{\squark{q}\antisquark{q}}+n_s^{\squark{q}\neutralino{1}}+n_s^{\neutralino{1}\neutralino{1}}\,, \label{eq: combine processes nsignal and xsec}
\end{equation}
where $n_s^{\squark{q}\antisquark{q}}$, $n_s^{\squark{q}\neutralino{1}}$ and $n_s^{\neutralino{1}\neutralino{1}}$ respectively denote the number of signal events expected for each of the three individual processes. We refer to \cref{appx: statistics} for more details.

\subsection{Combining the analyses}
\label{sec:combination}
The combination of the \glspl*{SR} of the four analyses requires to take into account their mutual correlations. For this purpose, we make use of \software{MA5} to generate a binary $n_\mathrm{events}\times n_\mathrm{SR}$ matrix containing one row per event and one column per \gls*{SR}. A specific element of this matrix is either 1 or 0, depending on whether a given event populates the corresponding \gls*{SR} or not. This `acceptance' matrix is next passed to \software{TACO}, which calculates the (Pearson) correlation coefficients between two \glspl*{SR} and generates a symmetric correlation matrix, \textit{i.e.} an $n_\mathrm{SR} \times n_\mathrm{SR}$ matrix. Two \glspl*{SR} are flagged as uncorrelated if their correlation coefficient is smaller than \SI{0.01}{}, else their correlation is stored. Moreover, the correlation coefficients relating a signal region of an ATLAS analysis to a signal region of a CMS analysis are manually set to zero, because they rely on distinct data sets. This is however an optimistic approximation, as the two experiments may share some systematic uncertainties. In this case, any non-zero correlation would reduce the combined exclusion limit obtained~\cite{Araz:2023bwx}.

The correlation matrix is then passed to the \software{PathFinder} software so that it could determine the most constraining combination of uncorrelated \glspl*{SR}. This is done by assigning to each \gls*{SR} a weight equal to the logarithm of the likelihood-ratios between the \gls*{SUSY} model tested and the background-only expectation, that is also proportional to the \glspl*{CL} exclusion. Since the likelihoods of multiple \glspl*{SR} combine multiplicatively, the log likelihood ratio associated with a subset of regions is simply given by the sum of their weights. The \software{PathFinder} package has been designed to identify the set of uncorrelated \glspl*{SR} with the highest accumulated weight. 

\begin{figure}
     \centering
     \begin{subfigure}[b]{0.49\textwidth}
         \centering
            \includegraphics[width=\textwidth]{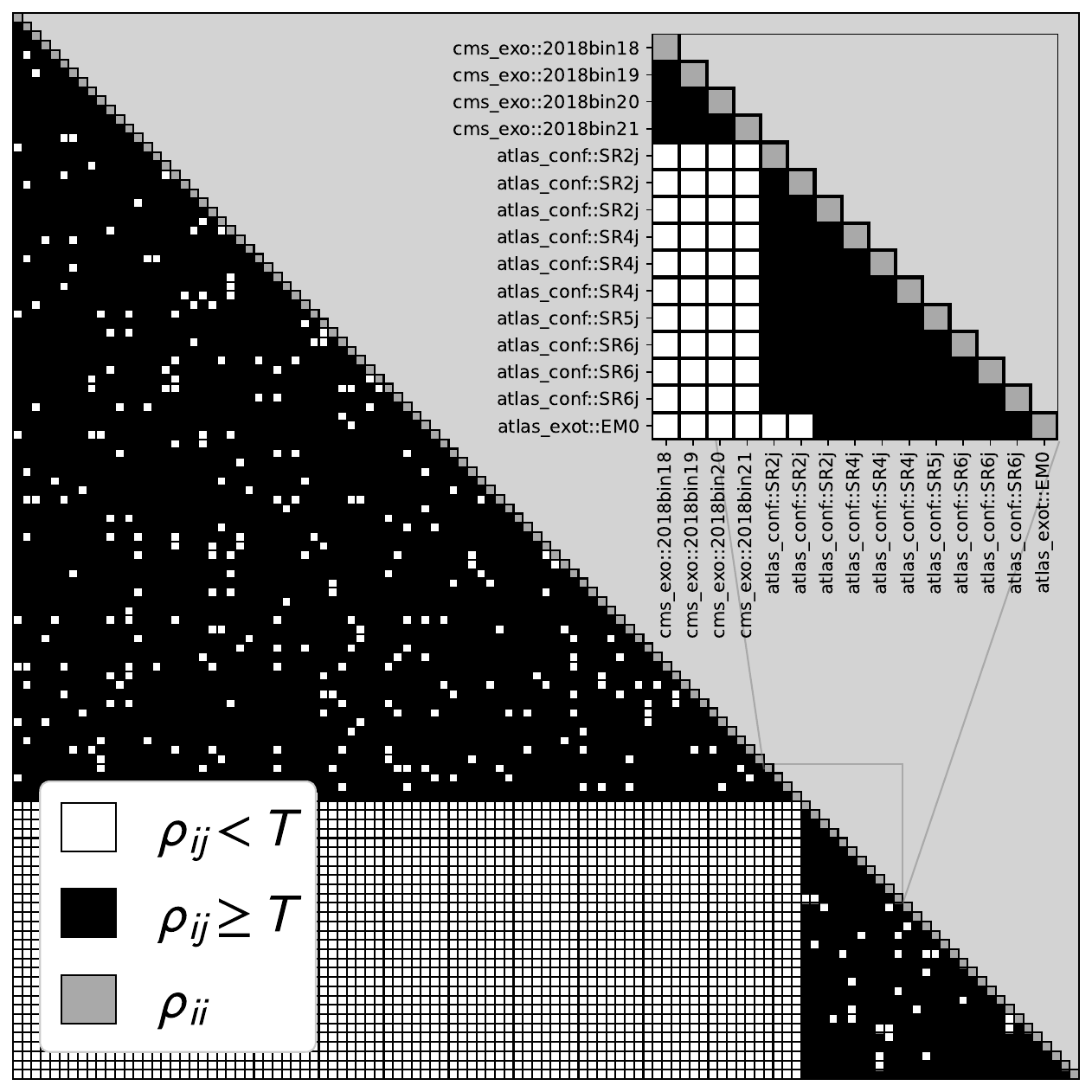}
            \caption{unsorted correlation matrix}
            \label{fig:pathfinder-cor_unsorted}
     \end{subfigure}
     \hfill
     \begin{subfigure}[b]{0.49\textwidth}
         \centering
         \includegraphics[width=\textwidth]{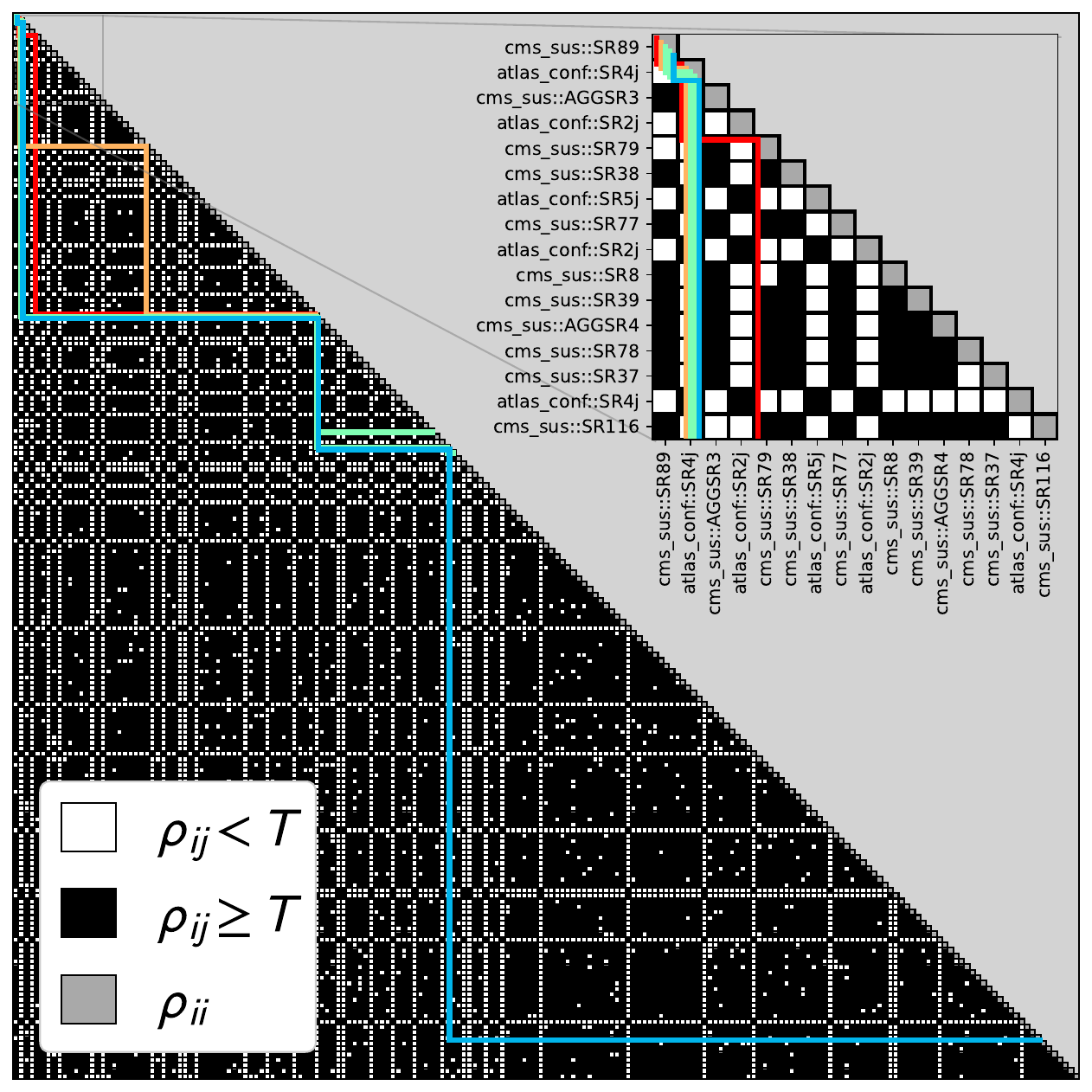}
        \caption{sorted correlation matrix}
         \label{fig:pathfinder-cor_sorted}
     \end{subfigure}
    \caption{Correlation matrix $\rho$ of a representative set of \glspl*{SR} of the four discussed analyses in case of squark pair production and a scenario with $m_{\neutralino{1}}=\SI{350}{\giga \electronvolt}$ and $m_{\squark{q}}=\SI{1000}{\giga \electronvolt}$. Black and white dots respectively encode a correlated and an uncorrelated pair of \glspl*{SR}. We display the correlation matrix with \glspl*{SR} unsorted (left), as well as after sorting them in a way such that \glspl*{SR} with the highest weight come first (right). The combinations yielding the best sensitivity are highlighted through coloured paths.}
    \label{fig:pathfinder}
\end{figure}

In \cref{fig:pathfinder}, we present various versions of the correlation matrix $\rho$ obtained for squark pair production, considering an illustrative scenario with $m_{\neutralino{1}}=\SI{350}{\giga\electronvolt}$ and $m_{\squark{q}}=\SI{1000}{\giga\electronvolt}$. First, we display the unsorted correlation matrix in~\cref{fig:pathfinder-cor_unsorted}, where the \glspl*{SR} are arranged based on the order of their declarations in the analysis implementations. Notably, a block-shaped white area is observed in the lower-left part of the figure. Each of the elements of this block represents the intersection of an ATLAS \gls*{SR} and a CMS \gls*{SR}, which are enforced to be uncorrelated due to their association with different experiments. After sorting the columns and rows of the matrix $\rho$ according to the calculated weight of its different elements, we obtain the structured matrix shown in~\cref{fig:pathfinder-cor_sorted}, with now lines of mostly only uncorrelated regions instead of a block. The zoomed-in insets of the figures present results for a subset of \glspl*{SR}, identified by their names on the axes. It is evident that the majority of combinations within one analysis are correlated, but some uncorrelated options still exist. 

Utilising the path-finding algorithm, we identify the five best paths that combine the different regions in the most sensitive manner. 
A given path contains all \glspl*{SR} where the corresponding coloured line intersects the diagonal, with the red line denoting the most constraining combination. 
As can be seen from the zoomed-in inset, this best combination includes the two topmost and therefore most constraining individual \glspl*{SR}, and the fifth most constraining \gls*{SR}. This demonstrates that combinations of \glspl*{SR} of the same analysis can be realised (here CMS-SUS-19-006). These three \glspl*{SR}, as well as any set of \glspl*{SR} lying on a specific found path, are uncorrelated with each other (which is highlighted by the black-and-white color coding).
This can be seen from explicitly looking at the correlations in the fifth row, showing that the fifth \gls*{SR} is uncorrelated with both the first and second \gls*{SR}. However, the most constraining combination does not necessarily have to include the most constraining \gls*{SR}, but can also start further down, depending on the calculated combined weights.

We hence demonstrate the existence of combinations of a few regions that have the potential to enhance sensitivity to the signal. This is discussed quantitatively in~\cref{sec:results}, after inputting the list of \glspl*{SR} corresponding to the most constraining path into the software tool \software{SPEY} to calculate exclusions. We note that to combine the three sub-processes considered, we need to construct the corresponding acceptance matrix with all generated events included. This is achieved through the merging of the three individual acceptance matrices along the `\emph{event} axis'.

\section{Results}\label{sec:results}

\begin{figure}
     \centering
     \begin{subfigure}[b]{0.49\textwidth}
         \centering
         \includegraphics[width=\textwidth]{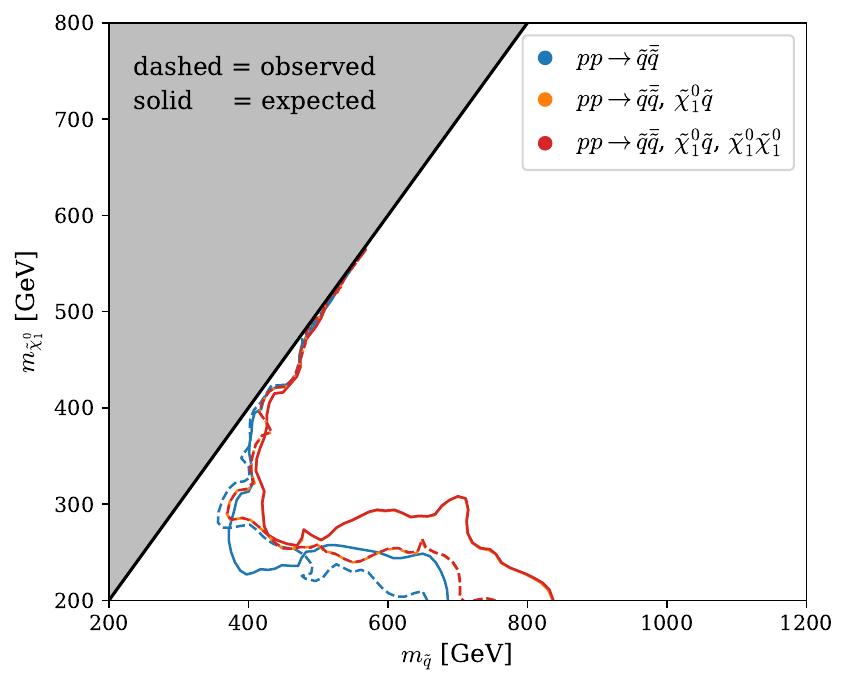}
         \caption{ATLAS-EXOT-2018-06}\vspace*{0.6cm}
     \end{subfigure}
     \hfill
     \begin{subfigure}[b]{0.49\textwidth}
         \centering
         \includegraphics[width=\textwidth]{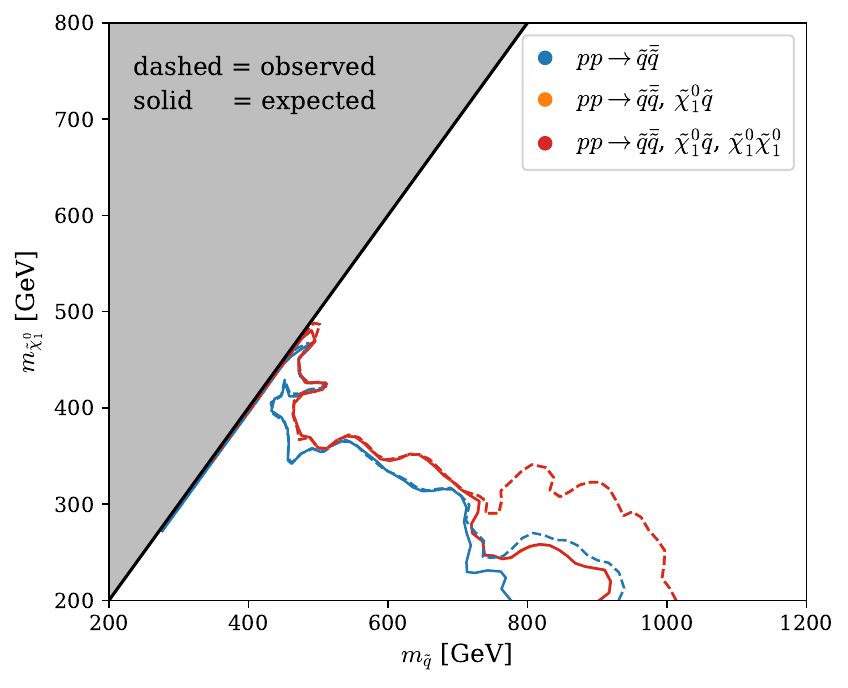}
         \caption{ATLAS-CONF-2019-040}\vspace*{0.6cm}
     \end{subfigure}
     \begin{subfigure}[b]{0.49\textwidth}
         \centering
         \includegraphics[width=\textwidth]{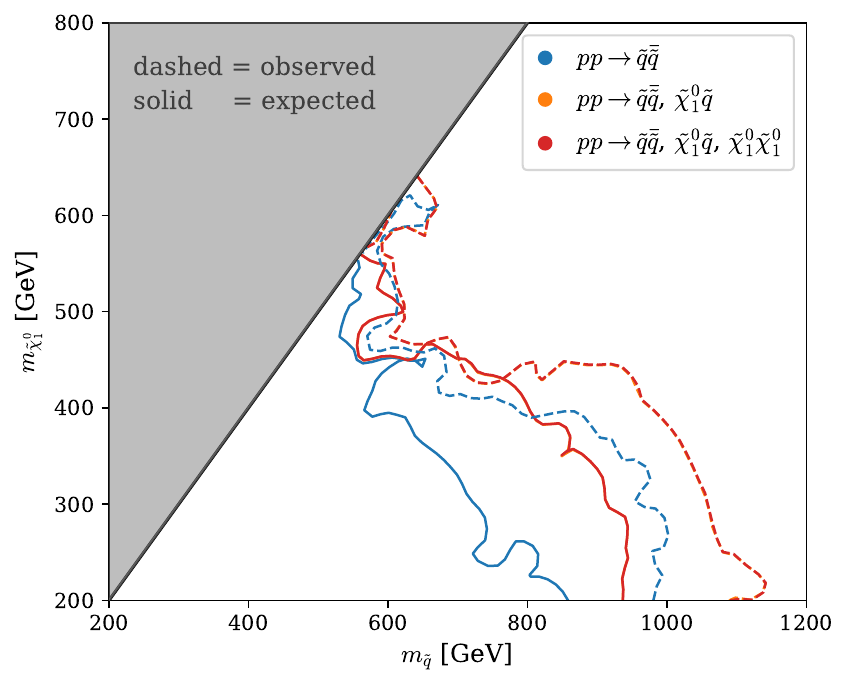}
         \caption{CMS-SUS-19-006}
     \end{subfigure}
     \hfill
     \begin{subfigure}[b]{0.49\textwidth}
         \centering
         \includegraphics[width=\textwidth]{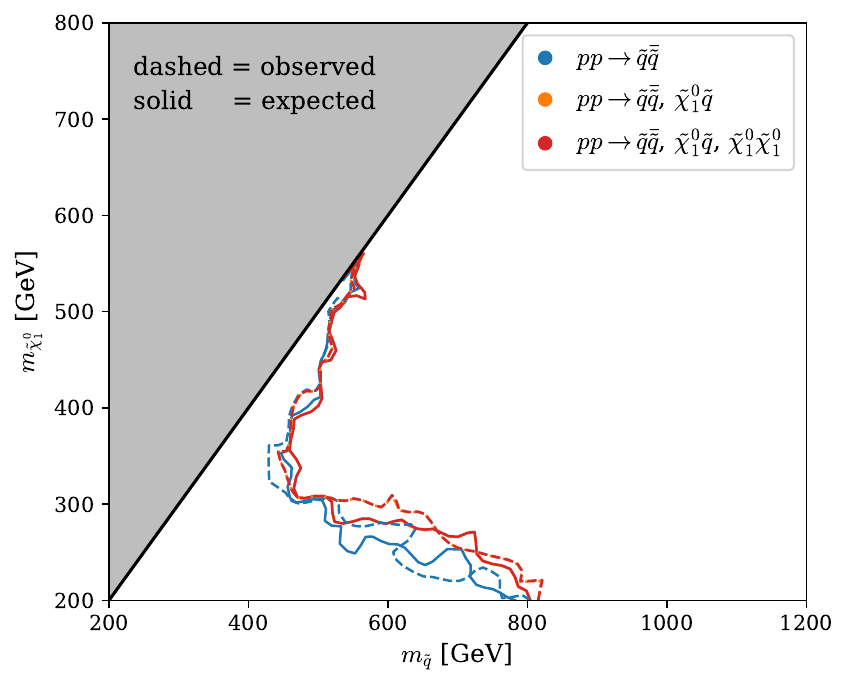}
         \caption{CMS-EXO-20-004}
     \end{subfigure}
    \caption{Exclusion contours at \SI{95}{\percent} CL on the considered simplified model, presented in the squark and neutralino mass plane. Bounds have been obtained by the reinterpretation of the results of the ATLAS-EXOT-2018-06 (top left), ATLAS-CONF-2019-040 (top right), CMS-SUS-19-006 (bottom left) and CMS-EXO-20-004 (bottom right) analyses. We consider three scenarios for signal modelling: a signal involving only squark-pair production (blue), both squark-pair and associated neutralino-squark production (orange), and all three production sub-processes (red). Expected (solid) and observed (dashed) bounds are derived from the most constraining \gls*{SR} in each specific analysis.}
    \label{fig:exclusion_individual_analyses}
\end{figure}

In \cref{fig:exclusion_individual_analyses} (as well as in the tables provided in \cref{app: tables}), we focus on the bounds obtained from each of the four considered analyses individually. Expected (solid lines) and observed (dashed lines) \SI{95}{\percent} \gls*{CL} exclusion limits are presented in the $(m_{\squark{q}}, m_{\neutralino{1}})$ mass plane, and we compute them based on signal predictions determined as outlined in \cref{sec:analysis}. The exclusion contours are computed by interpolating between the considered mass points (see \cref{app: tables}). Their smoothness and regularity is therefore a result of both the parameter space coverage regarding our simplified model of the considered analysis and the quality of the interpolation. Each figure showcases sensitivity contours distinguished by their colour coding. We compare bounds that are determined from a signal involving only strong production of a pair of squarks, each decaying into a neutralino and jets (blue). Such a signal definition matches how \gls*{SUSY} signals are simulated in experimental analyses up to now. Next, we examine the variation of the bounds when contributions from associated neutralino-squark production (orange) and neutralino-pair production (red) are included. We recall that these limits are derived by considering solely the most sensitive among all \glspl*{SR} within each analysis.

The `full' bounds, incorporating all three sub-processes (red contours), largely coincide with the bounds derived from the combination of squark-pair and associated neutralino-squark production only (orange contours). This suggests that neutralino pair production has a marginal impact on the exclusions computed for a given mass spectrum. While the strong channel contributes dominantly, associated production has a non-negligible impact and therefore allows for a strengthening of the limits. This is particularly evident in the parameter space regions where the squark mass is large, and where, as already discussed in \cref{sec:theory} regarding cross sections, squark-pair production begins to be phase-space suppressed. For example, at $m_{\neutralino{1}}=\SI{250}{\giga \electronvolt}$, both observed and expected exclusion limits exhibit gains in squark mass of approximately \SI{10}{\giga \electronvolt} for the CMS-EXO-20-004 analysis, and up to about \SI{100}{\giga \electronvolt} for the other three analyses. These results reinforce the need for better and more accurate signal modelling in all existing LHC searches for \gls*{SUSY}, whenever it is computationally achievable.

Moreover, the four analyses demonstrate sensitivity to different regions in the parameter space, hinting at the potential to enhance overall sensitivity through their combination. Notably, the ATLAS-EXOT-2018-06 and CMS-SUS-19-006 analyses show increased sensitivity in the vicinity of the diagonal of the mass plane, where the squark-neutralino spectrum is compressed. In such scenarios, neutralinos emerging from squark decays carry almost all the squark energy so that the associated jets have low energy. Consequently, the overall jet multiplicity in the signal events is lower than in split mass configurations where highly-energetic jets could originate from squark decays. Furthermore, the CMS-SUS-19-006 analysis yields the strongest exclusion limits among all four analyses considered. Unlike the others, there is no requirement on the presence of a specific very highly energetic jet. Instead, it imposes constraints on global hadronic activity through the $H_T$ variable defined in~\cref{eq:globalobs}, ensuring that it surpasses a certain threshold. Consequently, this analysis allows for several jets with smaller transverse momentum to collectively fulfil the hadronic activity requirements, rather than relying on a single jet embedding the bulk of it. The dominant squark pair production channel typically leads to the production of at least two hard jets from squark decays, with the possibility of additional hadronic activity from QCD radiation, thereby meeting the analysis thresholds outlined in \cref{tab:analyses cuts} without needing to rely on the presence of a very hard initial-state radiation. Thus, at this point, we conclude that analyses that are more inclusive in terms of jet energy and multiplicity are preferable to probe the considered MSSM-inspired simplified model.

\begin{figure}
     \centering
     \begin{subfigure}[b]{0.49\textwidth}
         \centering
         \includegraphics[width=\textwidth]{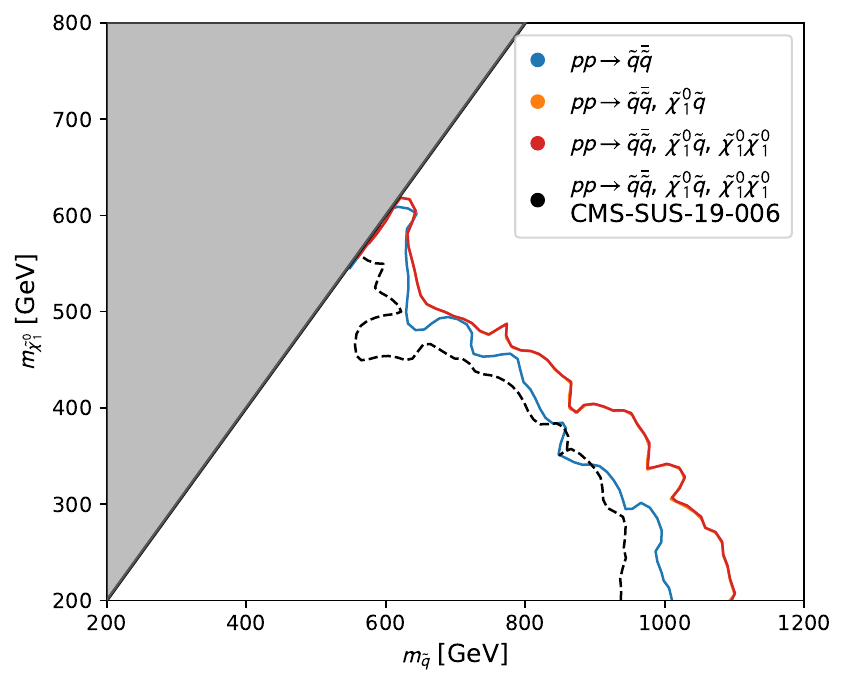}
         \caption{expected}
         \label{fig:exclusion_atlas_exot}
     \end{subfigure}
     \hfill
     \begin{subfigure}[b]{0.49\textwidth}
         \centering
         \includegraphics[width=\textwidth]{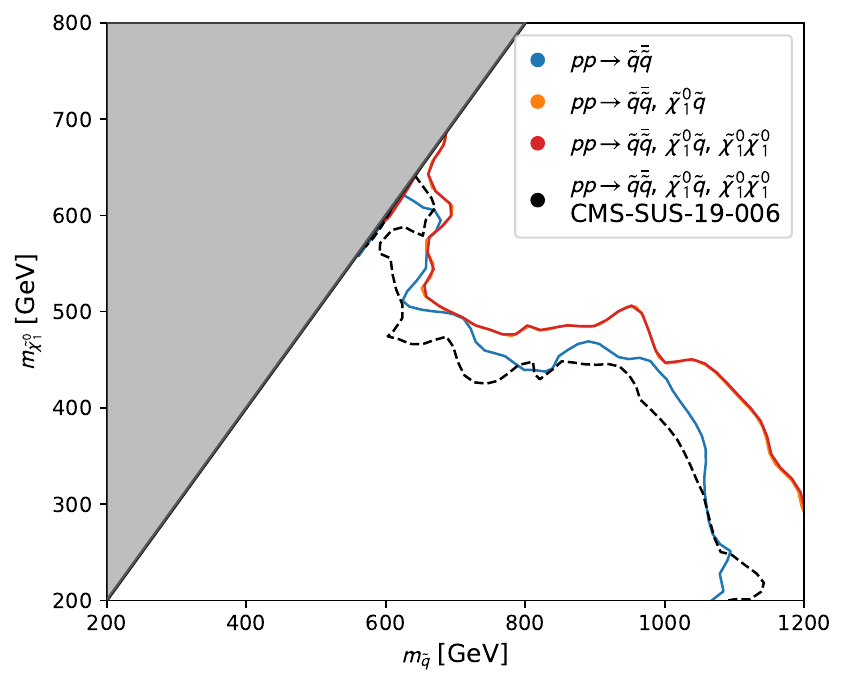}
         \caption{observed }
         \label{Expected (left) and observ}
     \end{subfigure}
    \caption{Expected (left) and observed (right) exclusion contours at \SI{95}{\percent} CL on the considered simplified model, presented in the squark and neutralino mass plane. Bounds have been derived from the most sensitive combination of uncorrelated \glspl*{SR} of the ATLAS-EXOT-2018-06, ATLAS-CONF-2019-040, CMS-SUS-19-006 and CMS-EXO-20-004 analyses. We consider three scenario for signal modelling: a signal involving only squark-pair production (blue), both squark-pair and associated neutralino-squark production (orange), and all three production sub-processes (red). For comparison, the exclusion contour determined from the most sensitive \gls*{SR} of the CMS-SUS-19-006 analysis is also shown (dashed, black).}
    \label{fig:exclusion_combined}
\end{figure}

The exclusion limits obtained by combining uncorrelated \glspl*{SR} from the four different analyses are depicted in \cref{fig:exclusion_combined} (as well as in the tables of \cref{app: tables}). Bounds are computed from the most sensitive combination of \glspl*{SR} determined by the \software{PathFinder} package, as detailed in \cref{sec:combination}. We analyse signal scenarios involving squark-pair production only (blue), and assess the impact of adding contributions from associated neutralino-squark production (orange) and neutralino-pair production (red). As above, neutralino-pair production has no discernible effect, as evidenced by the superposition of the red and orange contours. The comparison of the bounds computed from a signal comprising only squark-pair production, as currently modelled in experimental LHC analyses, to those derived from the full squark and neutralino signal, underscores once again the significant potential improvement achievable through better signal modelling including all contributing sub-processes, especially when the spectrum is not too compressed. 

We compare these findings with limits derived from a signal including all three sub-processes, and utilising the most sensitive \gls*{SR} of the CMS-SUS-19-006 analysis (dashed contour). This comparison is motivated by the fact that the CMS-SUS-19-006 analysis has consistently emerged as the most constraining analysis (see \cref{fig:exclusion_individual_analyses}). Combining multiple uncorrelated \glspl*{SR} of different analyses results in an additional gain in sensitivity compared to using the best \gls*{SR} from individual analyses to derive bounds, the gain being more pronounced when the spectrum is more split. Bounds on squark masses typically increase by approximately \SI{100}{\giga \electronvolt} for both expected and observed limits, reaching up to \SI{200}{\giga \electronvolt} in certain regions of the parameter space. Notably, employing a conservative modelling of the \gls*{SUSY} signal including only squark-pair production together with the best combination of uncorrelated regions (blue contours) yields increased sensitivity compared to relying on the best signal modelling possible but without combining any \gls*{SR}. Thus, while improving signal modelling is important, correlating findings from different analyses or sub-analyses is equally, and possibly more, essential.

\begin{figure}
     \centering
     \begin{subfigure}[b]{0.49\textwidth}
         \centering
         \includegraphics[width=\textwidth]{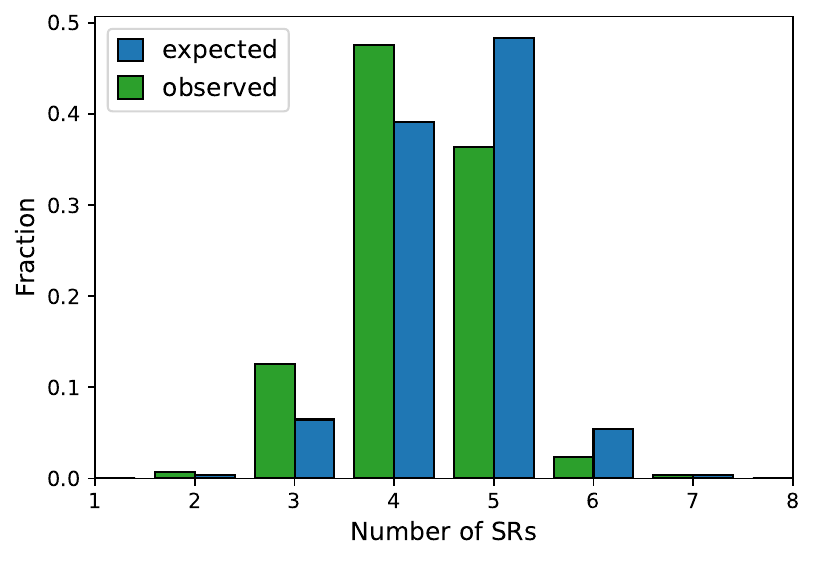}
         \caption{Number of combined \glspl*{SR} }
         \label{fig:stats_num_of_SRs}
     \end{subfigure}
     \hfill
     \begin{subfigure}[b]{0.49\textwidth}
         \centering
         \includegraphics[width=\textwidth]{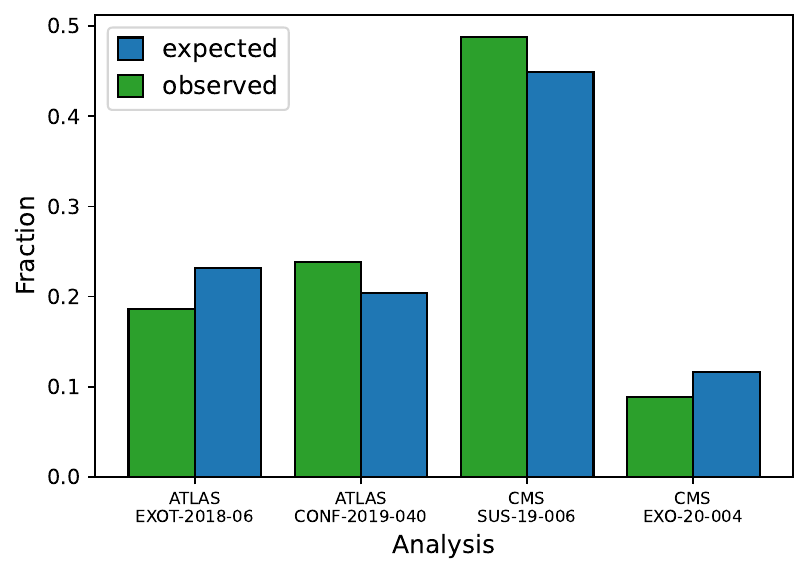}
         \caption{Share of the different analyses}
         \label{fig:stats_num_of_analyses}
     \end{subfigure}
    \caption{Information on the combination of \glspl*{SR} across the parameter space. For each scanned mass point, we report the number of \glspl*{SR} combined (left), and the proportion of \glspl*{SR} originating from each individual analysis (right). The results are normalised to the total number of scanned points in the parameter space.}
    \label{fig:stats}
\end{figure}

To understand the details of the \gls*{SR} combination process for the considered signal, we present in \cref{fig:stats} the number of \glspl*{SR} combined according to the results of the \software{Pathfinder} algorithm (left panel). Additionally, we indicate, for each specific combination, the proportion of \glspl*{SR} originating from each individual analysis (right panel). In both cases, we normalise the results to the total number of scanned points in the parameter space. On average, four to five different \glspl*{SR} are combined in more than 90\% of the cases, with the CMS-SUS-19-006 analysis contributing the most to this combination. This reflects the fact that this analysis is individually the most constraining, being more inclusive in its selection criteria (see also \cref{tab:analyses cuts}). However, it has a leading and determining impact only in half of the cases, highlighting the importance of exploiting information from other analyses as well.
The average number of combined \glspl*{SR} that belong to the same analysis can be approximated by multiplying the average path length deduced from \cref{fig:stats_num_of_SRs} with the relevant analysis fraction shown in \cref{fig:stats_num_of_analyses}. This yields roughly 2.5 for the CMS-SUS-19-006 analysis, 1.0 for each of the considered ATLAS analyses, and 0.5 for the CMS-EXO-20-004 search.

While other analyses are generally less constraining than the CMS-SUS-19-006 analysis, they still substantially contribute to the combination. This is particularly true for the ATLAS-EXOT-2018-06 and ATLAS-CONF-2019-040 analyses, as ATLAS and CMS searches are naturally considered uncorrelated. Furthermore, although we might expect a reduced contribution from the monojet CMS-EXO-20-004 analysis due to its smaller relevance (as displayed in \cref{fig:exclusion_individual_analyses}) and a likely stronger correlation with the CMS-SUS-19-006 analysis, its effects cannot be neglected. Therefore, we advocate for combining analyses whenever possible to ensure the best possible coverage of the parameter space with available data.

\section{Conclusion}
\label{sec:concl}
In this project, we focused on a simplified supersymmetric scenario inspired by the \gls*{MSSM}, where all superpartners are decoupled except for one squark flavour and a bino-like neutralino. Our objective was to demonstrate the gain in exclusion power achieved by combining uncorrelated \glspl*{SR} from different LHC searches for \gls*{SUSY}. Additionally, we aimed to illustrate the importance of improved signal modelling inherent to new physics setups in which several processes contribute. Our analysis encompasses four LHC searches targeting multiple jets and missing transverse energy, specifically the ATLAS-EXOT-2018-06, ATLAS-CONF-2019-040, CMS-SUS-19-006 and CMS-EXO-20-004 analyses. We considered a signal comprising squark-pair production, associated squark-neutralino production, as well as neutralino-pair production, that we modelled using state-of-the-art Monte Carlo simulation techniques including multi-partonic matrix element merging and the most precise predictions to date for the corresponding cross sections.

We began by examining a signal incorporating squark-pair production only, consistent with what is achieved in Monte Carlo simulations relevant for current LHC searches. Our investigation revealed that the addition of associated squark-neutralino production significantly enhanced the exclusion power, corroborating recent findings. This effect is particularly pronounced when the \gls*{SUSY} spectrum is split and for neutralino masses below approximately \SI{400}{\giga \electronvolt}. Furthermore, we observed that neutralino-pair production has minimal impact on the currently probed parameter space regions with the available data set. However, this may change in the future due to upcoming stronger squark and neutralino bounds, and the associated phase-space suppression of squark-pair and neutralino-squark production.

Combining uncorrelated \glspl*{SR} from different analyses further extends the reach of the searches considered. Notably, squark and neutralino mass bounds exceed those derived from individual analyses by approximately \SI{100}{\giga \electronvolt} to \SI{200}{\giga \electronvolt}. These results highlight the critical importance of combining \glspl*{SR} from different analyses, ideally within LHC collaborations when feasible, but also through innovative and approximate methods like the recently developed \software{TACO} approach utilised in this study and that could be employed outside the collaborations.

\acknowledgments
We thank Jack~Araz for providing us with useful inputs regarding the \software{SPEY} package, as well as Humberto~Reyes-Gonzalez and Jamie~Yellen for their help with the \software{Taco} and the \software{PathFinder} software. We also thank Sabine Kraml for reading and commenting on the manuscript. BF acknowledges support from grant ANR-21-CE31-0013 (project DMwithLLPatLHC) from the French \emph{Agence Nationale de la Recherche} (ANR). Work in Münster is supported by the BMBF through project 05P21PMCAA and DFG through GRK 2149 and SFB 1225 “Isoquant,” project-id 273811115.
JF acknowledges financial support from ICSC~– Centro Nazionale di Ricerca in High Performance Computing, Big Data and Quantum Computing, funded by European Union – NextGenerationEU.
Calculations (or parts of them) for this publication were performed on the HPC cluster PALMA II of the University of Münster, subsidised by the DFG (INST 211/667-1).

\appendix
\section{Event generation and multi-partonic matrix element merging}
\label{appx: event generation}
In order to get an accurate description of the hadronic activity in events with a final state featuring a high jet multiplicity after parton showering, it is essential to rely on the combination of hard-scattering matrix elements possibly featuring additional QCD radiation. Consistency in this combination is crucial to avoid double counting the QCD emissions generated from the matrix elements (describing exclusive final states with a fixed number of jets) and from parton showers (describing inclusive final states in the number of jets). We achieve this by following the CKKW-L  merging prescription~\cite{Catani:2001cc, Lonnblad:2001iq, Lonnblad:2011xx}. In practice, the algorithm starts from a hard-scattering event and generates a sequence of dipole emissions resembling the structure emerging from the showered event. If the configuration of any of the dipole emissions falls into the phase space region that should be described by the matrix element, the event is rejected. This ensures that the hardest emissions are always described by matrix elements and the softer ones by parton showers. Analogously, this results in re-weighting the event by the associated Sudakov form factor (\textit{i.e.}\ the probability that there is no further emission once we start from the matrix element configuration). Each dipole emission in the sequence is thus accepted as long as its scale does not exceed a predefined merging scale $Q_{\text{MS}}$, that describes the  transition between the regime in which QCD radiation is described by matrix elements and that in which it is described by parton showers.

\Gls*{DJR} distributions play an important role in validating the choice of parameters relevant for the merging prescription, particularly the value of $Q_{\text{MS}}$ to which they are highly sensitive. A suitable choice ensures a smooth transition between an event topology with $n-1$ final-state hard jets and a topology with $n$ final-state hard jets, without any bumps, peaks or abrupt transitions in the \gls*{DJR} spectra. However, default settings in \software{MG5aMC} often result in bumps indicative of resonances for the considered signal. These bumps stem from certain diagrams exhibiting the exchange of intermediate $s$-channel off-shell squarks that see their off-shell nature modified by QCD radiation. To address this issue, the \texttt{bwcutoff} parameter in the \software{MG5aMC} configuration card has to be increased from \SI{15}{\nothing} to, for example, \SI{35}{\nothing}. This adjustment implies that squarks are flagged as on-shell in a wider region around their pole mass, which further impacts cutting out their contributions in processes featuring neutralinos in the final state (see \cref{sec:theory}).

\begin{figure}
    \centering
    \begin{subfigure}[b]{\textwidth}
        \includegraphics[width=\textwidth]{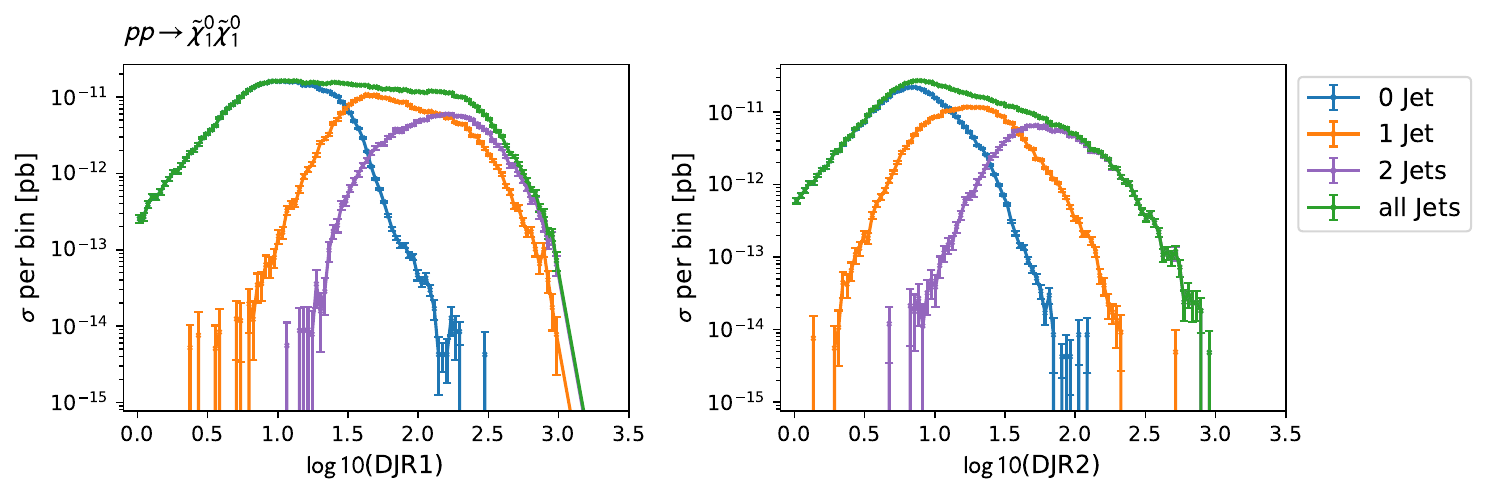}
    \end{subfigure}
    \begin{subfigure}[b]{\textwidth}
        \includegraphics[width=\textwidth]{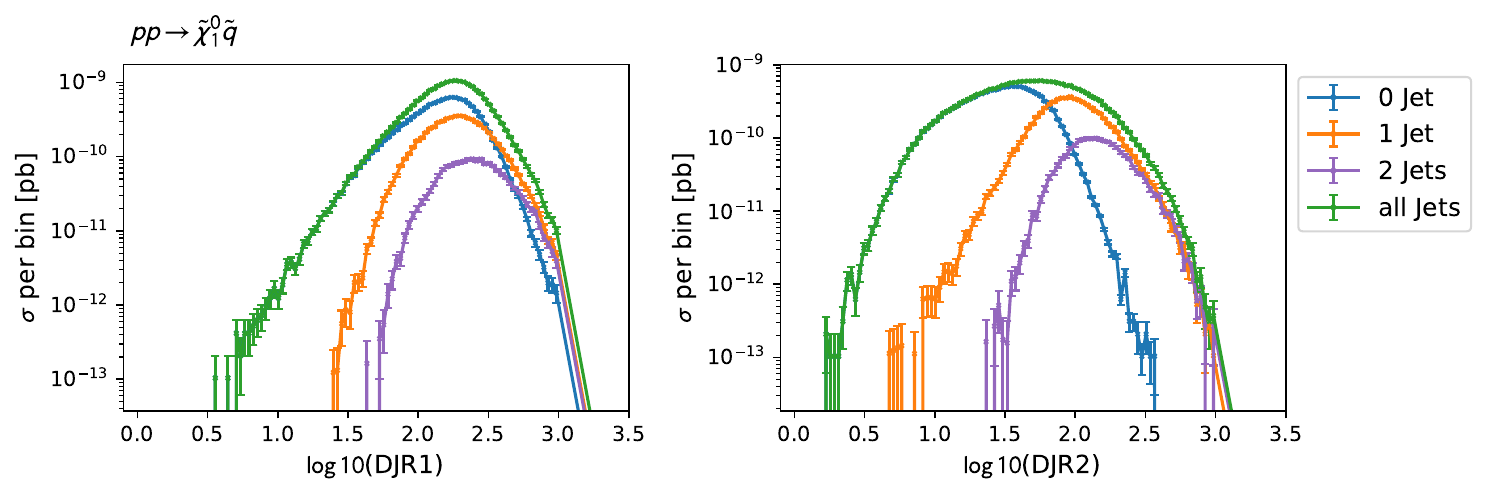}
    \end{subfigure}
    \begin{subfigure}[b]{\textwidth}
        \includegraphics[width=\textwidth]{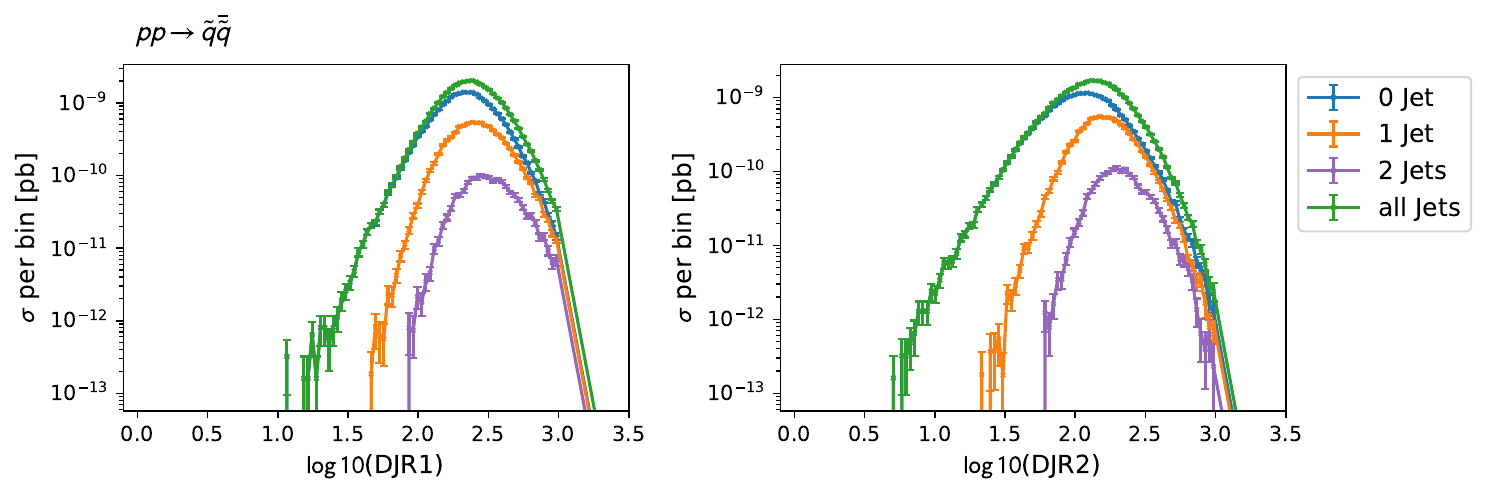}
    \end{subfigure}
    \caption{\gls*{DJR} distributions for the three different processes considered, and a scenario defined by $m_{\neutralino{1}}=\SI{200}{\giga \electronvolt}$ and $m_{\squark{q}}=\SI{600}{\giga \electronvolt}$.}
    \label{fig:DJR_200_600}
\end{figure}

We validated the merging prescription for a variety of different mass combinations, that we illustrate for the case $m_{\neutralino{1}}=\SI{200}{\giga \electronvolt}$ and $m_{\squark{q}}=\SI{600}{\giga \electronvolt}$ in~\cref{fig:DJR_200_600}. We present, for neutralino-pair (top row), squark-neutralino (middle row) and squark-pair (bottom row) production, total \gls*{DJR} spectra (green), together with the individual contributions emerging from matrix elements featuring no extra jet (blue), one extra jet (orange) and two extra jets (purple). For completeness, we provide below the associated \software{MG5aMC} commands that we use to test associated neutralino-squark production (the commands for the other channels being similar):
\begin{verbatim}
    define squa = ur ur~
    generate p p > n1 ur $squa, ur > u n1 @1
    add process p p > n1 ur~ $squa, ur~ > u~ n1 @1
    add process p p > n1 ur j $squa, ur > u n1 @2
    add process p p > n1 ur~ j $squa, ur~ > u~ n1 @2
    add process p p > n1 ur  j j $squa, ur > u n1 @3
    add process p p > n1 ur~ j j $squa, ur~ > u~ n1 @3
    output
    launch
    
    shower = Pythia8 
    set pdlabel = lhapdf 
    set lhaid = 27000 
    set ebeam1 = 6500.0 
    set ebeam2 = 6500.0 
    set use_syst = False 
    set nevents = 500000 
    set mass mneu1 200.0 
    set mass msu4 600.0 
    set mass msd1 30000 
    set mass msd2 30000 
    ...                         #all other SUSY masses decoupled            
    set mass msn2 30000 
    set mass msn3 30000 
    set wsu4 auto 
    set rnn1x1 1 
    set rnn2x2 1 
    set rnn1x2 0 
    set rnn1x3 0 
    ...                         #all other entries set to 0
    set rnn4x3 0 
    set rnn4x4 0 
    set Merging:nJetMax 2 
    set ktdurham = -1 
    set Merging:mayRemoveDecayProducts = on 
    set Merging:doPTLundMerging=on 
    set Merging:Process=pp>n1,ur,ur~ 
    set ptlund = 100.0               #ptlund=1/8 (m_squark+m_neutralino)
    set bwcutoff = 35.0
\end{verbatim}
We can note that we have explicitly chosen the merging scale to one quarter of the \gls*{SUSY} hard scale as given in \cref{eq. merging scale}.

\section{Statistical analysis}
\label{appx: statistics}
Within the \software{SPEY} framework~\cite{Araz:2023bwx}, the likelihood function for signal exclusion is expressed as
\begin{align}
    \llh(\mu, \theta)&= \text{Poiss}\left(n_{\text{obs}}| \mu n_s+ n_b +\theta \cdot \Delta n_b\right) \text{Gauss}\left(\theta\right|0, 1)\,. \label{eq. likelihood basic}
\end{align}
Here, $\mu$ represents the signal strength, indicating in our case the presence ($\mu=1$) or absence ($\mu=0$) of a \gls*{SUSY} signal. The parameter $\theta$ serves as a single nuisance parameter and captures how results can deviate from the \gls*{SM} expectation in terms of standard deviations. In this formulation, the likelihood function comprises two components. The Poisson-distributed factor accounts for the observed number of events given the expected contributions from both signal $n_s$ and background $n_b$, along with the impact of the nuisance parameter that multiplies the uncertainty in the background estimation denoted by $\Delta n_b$. Furthermore, a unit Gaussian distribution is used for the constraint factor to model uncertainties in the background estimation.

The \gls*{SUSY} signal hypothesis is tested using the test statistic $\tilde{q}_\mu$~\cite{Cowan:2010js}, defined by
\begin{align}
    \tilde{q}_{\mu}=
\begin{cases}
    -2\log\frac{\llh(\mu,\theta_\mu)}{\llh(0,\theta_0)} & \text{if } \hat{\mu} < 0 , \\[.1cm]
    -2\log\frac{\llh(\mu,\theta_\mu)}{\llh(\hat{\mu},\hat{\theta})} & \text{if } 0\leq \hat{\mu} \leq \mu , \\[.1cm]
    0 & \text{if } \mu < \hat{\mu} ,
\end{cases}
\end{align}
where $\hat{\mu}$ and $\hat{\theta}$ denote the set of parameters that maximises the likelihood (representing thus the best fit to the data), and $\theta_\mu$ is the nuisance parameter value that maximises the likelihood for a given signal strength $\mu$. The corresponding $p$-value (allowing to deduce an exclusion confidence level \gls*{CL}) is therefore given by
\begin{equation}
    p_{_{\tilde{q}_{\mu}}}= \int_{-\infty}^{\sqrt{\tilde{q}_{\mu}}-\sqrt{\tilde{q}_{\mu,A}}} f(\tilde{q}_{\mu}^{\prime}, \mu)\, \mathrm{d}\tilde{q}_{\mu}^{\prime}, \label{eq. pvalue and CL}
\end{equation}
where the test statistics $\tilde{q}_{\mu,A}$ is obtained assuming Asimov data and $f(\tilde{q}_{\mu}^{\prime}, \mu)$ corresponds to the $\tilde{q}_{\mu}$ probability distribution. Consequently, an increasing incompatibility between an hypothesised signal strength $\mu$ and the observed one $\hat{\mu}$ would lead to a decreasing likelihood ratio, resulting in an increasing test statistic and a higher corresponding \gls*{CL}.

\begin{figure}
    \centering
    \includegraphics[width=0.5\textwidth]{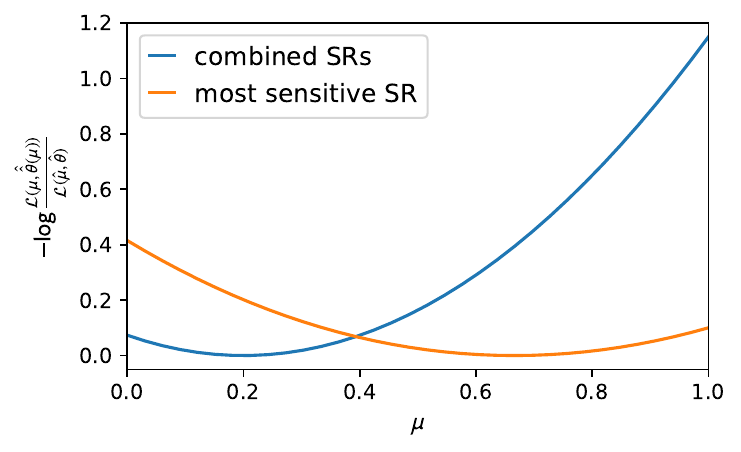}
    \caption{Negative logarithmic likelihood ratio as a function of the signal strength $\mu$ when using the most sensitive \gls*{SR} from all analyses (orange) and the most sensitive combination of \glspl*{SR} as returned by the \software{PathFinder} algorithm. We consider squark-pair production for $m_{\neutralino{1}}=\SI{250}{\giga \electronvolt}$ and $m_{\squark{q}}=\SI{1000}{\giga \electronvolt}$.}
    \label{fig:combined_likelihood}
\end{figure}
We compute both the background-only exclusion confidence level $\text{\gls*{CL}}_\text{b}$ (with $\mu=0$) and the signal-plus-background exclusion confidence level $\text{\gls*{CL}}_{\text{s}+\text{b}}$ (with $\mu=1$). The ratio $\text{\gls*{CL}}_\text{s}=\text{\gls*{CL}}_{\text{s}+\text{b}}/\text{\gls*{CL}}_\text{b}$ then yields the observed $p$-value. We obtain the expected $p$-value (referred to as the `\textit{a posteriori}' $p$-value in \software{SPEY}) by replacing the observed number of events with the expected number of \gls*{SM} events based on background predictions and uncertainties.

The effect of combining uncorrelated \glspl*{SR} is evaluated by multiplying their likelihoods,
\begin{equation}
    \llh_{\text{combined}}(\mu)=\prod \limits_{i \in \text{SRs}} \llh_i(\mu,\theta_i)\, . \label{eq. combined likelihood}
\end{equation}
This enables the calculation of \glspl*{llr} by simple addition, which justifies the weighting assigned by the \software{PathFinder} algorithm as discussed in \cref{sec:combination}. The impact of combining \glspl*{SR} on the likelihood is illustrated in \cref{fig:combined_likelihood} for a scenario with $m_{\neutralino{1}}=\SI{250}{\giga \electronvolt}$ and $m_{\squark{q}}=\SI{1000}{\giga \electronvolt}$, considering a signal comprising only squark-pair production. We compare the \gls*{llr} associated with the most sensitive \gls*{SR} from all analyses considered (orange) to that obtained with the most sensitive combination of \glspl*{SR} determined by the \software{PathFinder} algorithm (blue). For $\mu=1$ (the \gls*{BSM} signal hypothesis), the combined \gls*{llr} exceeds that of the single most sensitive \gls*{SR}, indicating a greater discrepancy between data and the signal-plus-background hypothesis and leading to a higher exclusion \gls*{CL}. Similarly, the shift of the minimum of the \gls*{llr} as a function of $\mu$ towards lower $\mu$ values suggests that the data is more consistent with the \gls*{SM} hypothesis ($\mu=0$).

\newpage

\section{Tables}\label{app: tables}
\begin{table}[h]
    \centering
    \tiny{}
\begin{tabular}{c|lllll|lllll|lllll}
{$m_{\neutralino{1}}$/$m_{\squark{q}}$} & \multicolumn{5}{c}{$\tilde{q}\tilde{q}$} & \multicolumn{5}{c}{$\tilde{q}\tilde{q}-\tilde{\chi}\tilde{q}$} & \multicolumn{5}{c}{$\tilde{q}\tilde{q}-\tilde{\chi}\tilde{q}-\tilde{\chi}\tilde{\chi}$} \\[.2cm]
{GeV} & AC & AE & CS & CE & \textit{cmb}. & AC & AE & CS & CE & \textit{cmb}.& AC & AE & CS & CE & \textit{cmb}.\\
\midrule
200/900  &       0.88 &       0.68 &    0.92 &    0.61 &     0.99 &       0.95 &       0.86 &    0.99 &    0.79 &      1.0 &       0.95 &       0.86 &    0.99 &    0.79 &      1.0 \\
200/1000 &       0.78 &       0.54 &    0.78 &    0.47 &     0.97 &       0.89 &       0.72 &    0.86 &    0.59 &     0.99 &       0.89 &       0.73 &    0.86 &    0.59 &     0.99 \\
200/1200 &       0.53 &       0.27 &    0.59 &    0.24 &     0.73 &       0.67 &       0.45 &    0.71 &    0.39 &     0.91 &       0.68 &       0.45 &    0.72 &     0.4 &     0.91 \\
200/1300 &       0.46 &       0.26 &     0.5 &    0.25 &     0.66 &       0.59 &       0.42 &    0.66 &    0.43 &     0.84 &       0.59 &       0.42 &    0.66 &    0.43 &     0.85 \\
250/750  &        0.9 &       0.85 &    0.99 &     0.8 &      1.0 &       0.94 &       0.95 &     1.0 &    0.91 &      1.0 &       0.94 &       0.95 &     1.0 &    0.91 &      1.0 \\
250/850  &        0.9 &       0.74 &    0.98 &    0.79 &      1.0 &       0.95 &        0.9 &    0.99 &     0.9 &      1.0 &       0.95 &        0.9 &    0.99 &     0.9 &      1.0 \\
250/940  &       0.82 &       0.56 &    0.96 &    0.57 &     0.99 &       0.91 &       0.77 &    0.96 &    0.73 &      1.0 &       0.91 &       0.78 &    0.96 &    0.73 &      1.0 \\
250/950  &       0.79 &       0.58 &    0.75 &    0.49 &     0.95 &       0.88 &       0.75 &    0.88 &    0.66 &     0.98 &       0.88 &       0.75 &    0.88 &    0.67 &     0.98 \\
250/1000 &       0.76 &       0.51 &    0.75 &    0.38 &     0.93 &       0.86 &       0.69 &    0.83 &    0.55 &     0.99 &       0.86 &        0.7 &    0.83 &    0.55 &     0.99 \\
250/1050 &       0.68 &       0.48 &    0.82 &     0.5 &     0.92 &       0.79 &       0.67 &    0.83 &    0.61 &     0.96 &       0.79 &       0.67 &    0.83 &    0.61 &     0.96 \\
250/1080 &       0.62 &       0.42 &    0.65 &    0.33 &     0.87 &       0.78 &        0.6 &    0.82 &    0.46 &     0.95 &       0.78 &        0.6 &    0.82 &    0.46 &     0.96 \\
250/1100 &       0.72 &       0.39 &    0.62 &     0.3 &     0.87 &       0.75 &       0.55 &    0.74 &    0.45 &     0.92 &       0.75 &       0.55 &    0.75 &    0.45 &     0.92 \\
300/750  &       0.84 &       0.79 &    0.99 &    0.73 &      1.0 &       0.93 &        0.9 &     1.0 &    0.86 &      1.0 &       0.93 &        0.9 &     1.0 &    0.87 &      1.0 \\
300/850  &       0.84 &       0.68 &    0.94 &     0.7 &      1.0 &       0.87 &       0.83 &    0.98 &    0.81 &      1.0 &       0.87 &       0.83 &    0.98 &    0.82 &      1.0 \\
300/900  &       0.83 &       0.61 &    0.96 &    0.72 &     0.99 &       0.91 &       0.78 &    0.99 &     0.8 &      1.0 &       0.91 &       0.78 &    0.99 &     0.8 &      1.0 \\
300/1000 &       0.68 &       0.48 &    0.77 &    0.41 &      0.9 &       0.78 &       0.66 &    0.84 &    0.56 &     0.95 &       0.78 &       0.66 &    0.84 &    0.57 &     0.96 \\
300/1200 &       0.46 &       0.27 &    0.53 &    0.25 &     0.66 &       0.58 &       0.44 &    0.65 &    0.37 &     0.79 &       0.58 &       0.44 &    0.65 &    0.37 &     0.79 \\
350/650  &       0.87 &       0.68 &     1.0 &    0.67 &      1.0 &       0.95 &       0.81 &     1.0 &    0.77 &      1.0 &       0.95 &       0.81 &     1.0 &    0.77 &      1.0 \\
350/700  &       0.75 &        0.7 &    0.99 &     0.7 &      1.0 &       0.86 &       0.84 &     1.0 &    0.82 &      1.0 &       0.86 &       0.84 &     1.0 &    0.83 &      1.0 \\
350/800  &       0.77 &       0.59 &    0.93 &    0.64 &     0.99 &       0.84 &       0.75 &     1.0 &    0.78 &      1.0 &       0.84 &       0.75 &     1.0 &    0.78 &      1.0 \\
350/850  &       0.66 &       0.58 &    0.89 &    0.56 &     0.95 &       0.78 &       0.73 &    0.95 &    0.67 &     0.99 &       0.78 &       0.73 &    0.95 &    0.68 &     0.99 \\
350/900  &       0.69 &       0.49 &    0.82 &    0.48 &     0.93 &       0.81 &       0.69 &    0.92 &    0.63 &     0.99 &       0.81 &       0.69 &    0.92 &    0.63 &     0.99 \\
350/1000 &       0.64 &        0.4 &    0.64 &    0.45 &     0.89 &       0.74 &       0.56 &    0.73 &    0.53 &     0.94 &       0.74 &       0.57 &    0.73 &    0.53 &     0.94 \\
350/1100 &        0.5 &       0.31 &     0.6 &    0.45 &     0.82 &       0.59 &       0.42 &    0.65 &    0.51 &     0.84 &        0.6 &       0.43 &    0.65 &    0.51 &     0.84 \\
400/600  &       0.75 &       0.45 &     1.0 &    0.64 &      1.0 &       0.87 &       0.57 &     1.0 &    0.83 &      1.0 &       0.87 &       0.57 &     1.0 &    0.83 &      1.0 \\
400/750  &       0.83 &        0.5 &    0.97 &    0.68 &     0.99 &       0.87 &       0.67 &    0.99 &    0.74 &      1.0 &       0.87 &       0.68 &    0.99 &    0.74 &      1.0 \\
400/800  &       0.61 &       0.49 &     0.9 &     0.6 &     0.97 &       0.71 &       0.66 &    0.95 &    0.69 &     0.99 &       0.71 &       0.66 &    0.95 &    0.69 &     0.99 \\
400/850  &       0.59 &       0.49 &    0.85 &    0.45 &     0.92 &       0.68 &       0.63 &     0.9 &    0.57 &     0.96 &       0.68 &       0.63 &     0.9 &    0.58 &     0.96 \\
400/870  &       0.52 &       0.48 &    0.77 &    0.42 &     0.92 &       0.65 &       0.64 &    0.87 &    0.59 &     0.94 &       0.65 &       0.64 &    0.87 &    0.59 &     0.94 \\
400/920  &       0.59 &       0.41 &     0.8 &    0.41 &     0.93 &       0.72 &       0.57 &    0.89 &    0.58 &     0.95 &       0.72 &       0.57 &    0.89 &    0.59 &     0.95 \\
400/950  &       0.59 &        0.4 &    0.81 &     0.5 &     0.89 &       0.71 &       0.56 &    0.82 &    0.59 &     0.95 &       0.71 &       0.56 &    0.82 &    0.59 &     0.95 \\
400/1000 &       0.54 &       0.33 &    0.65 &    0.46 &     0.77 &       0.62 &       0.48 &    0.72 &     0.5 &      0.9 &       0.62 &       0.48 &    0.72 &     0.5 &      0.9 \\
400/1200 &       0.38 &       0.27 &    0.42 &    0.26 &     0.57 &       0.52 &       0.38 &    0.56 &    0.35 &     0.76 &       0.52 &       0.38 &    0.56 &    0.35 &     0.77 \\
450/550  &       0.17 &       0.29 &     1.0 &    0.52 &      1.0 &       0.17 &       0.51 &     1.0 &    0.87 &      1.0 &       0.17 &       0.51 &     1.0 &    0.87 &      1.0 \\
450/600  &       0.17 &       0.27 &     1.0 &    0.49 &      1.0 &       0.17 &       0.39 &     1.0 &    0.55 &      1.0 &       0.17 &        0.4 &     1.0 &    0.55 &      1.0 \\
450/650  &        0.6 &       0.25 &     1.0 &    0.42 &      1.0 &       0.71 &       0.36 &     1.0 &    0.49 &      1.0 &       0.71 &       0.36 &     1.0 &     0.5 &      1.0 \\
450/820  &       0.17 &       0.38 &     0.8 &    0.42 &     0.89 &       0.17 &       0.51 &    0.89 &     0.5 &     0.96 &       0.17 &       0.52 &    0.89 &     0.5 &     0.96 \\
450/850  &       0.17 &       0.37 &    0.84 &    0.42 &      0.9 &       0.17 &        0.5 &    0.85 &    0.49 &     0.93 &       0.17 &        0.5 &    0.85 &    0.49 &     0.93 \\
450/950  &       0.17 &       0.33 &    0.72 &    0.31 &     0.87 &       0.17 &       0.45 &    0.72 &    0.42 &     0.95 &       0.17 &       0.46 &    0.72 &    0.42 &     0.95 \\
500/520  &       0.79 &       0.69 &     1.0 &    0.94 &      1.0 &       0.84 &       0.77 &     1.0 &    0.95 &      1.0 &       0.84 &       0.77 &     1.0 &    0.95 &      1.0 \\
500/600  &       0.17 &        0.3 &    0.99 &     0.5 &      1.0 &       0.17 &       0.38 &    0.99 &    0.89 &      1.0 &       0.17 &       0.39 &    0.99 &    0.89 &      1.0 \\
500/650  &       0.17 &       0.17 &    0.79 &    0.61 &     0.92 &       0.17 &       0.26 &    0.89 &    0.64 &     0.98 &       0.17 &       0.26 &     0.9 &    0.64 &     0.98 \\
500/700  &       0.17 &       0.22 &    0.84 &    0.22 &     0.93 &       0.17 &        0.3 &    0.85 &    0.32 &     0.94 &       0.17 &        0.3 &    0.85 &    0.32 &     0.94 \\
500/750  &       0.17 &       0.23 &    0.86 &    0.35 &      0.6 &       0.17 &        0.3 &    0.87 &    0.41 &     0.67 &       0.17 &        0.3 &    0.87 &    0.41 &     0.67 \\
500/800  &       0.17 &       0.24 &    0.74 &    0.26 &     0.82 &       0.17 &       0.35 &    0.84 &    0.42 &     0.93 &       0.17 &       0.35 &    0.84 &    0.42 &     0.93 \\
500/850  &       0.48 &       0.28 &    0.68 &    0.29 &     0.79 &       0.52 &       0.38 &    0.75 &    0.36 &      0.9 &       0.52 &       0.38 &    0.75 &    0.36 &      0.9 \\
500/950  &       0.39 &       0.25 &    0.57 &    0.25 &     0.72 &       0.44 &       0.35 &    0.66 &    0.33 &     0.79 &       0.44 &       0.36 &    0.66 &    0.33 &     0.79 \\
550/700  &       0.41 &       0.14 &    0.64 &    0.33 &     0.72 &       0.46 &       0.23 &    0.71 &    0.37 &     0.87 &       0.46 &       0.23 &    0.71 &    0.37 &     0.87 \\
550/750  &       0.41 &       0.11 &    0.54 &     0.3 &     0.69 &       0.47 &       0.18 &    0.61 &    0.33 &     0.75 &       0.48 &       0.18 &    0.61 &    0.33 &     0.75 \\
550/800  &        0.5 &       0.17 &    0.72 &    0.19 &     0.81 &       0.53 &       0.24 &    0.78 &    0.29 &     0.89 &       0.53 &       0.24 &    0.78 &    0.29 &     0.89 \\
550/850  &       0.43 &       0.17 &    0.92 &    0.22 &     0.94 &       0.47 &       0.25 &    0.92 &    0.26 &     0.95 &       0.47 &       0.25 &    0.92 &    0.26 &     0.95 \\
550/900  &       0.31 &        0.2 &    0.56 &    0.18 &     0.65 &       0.34 &       0.28 &    0.59 &    0.24 &     0.76 &       0.34 &       0.28 &    0.59 &    0.24 &     0.76 \\
600/650  &       0.56 &        0.2 &    0.89 &    0.32 &     0.94 &        0.6 &       0.28 &    0.89 &    0.36 &     0.93 &        0.6 &       0.28 &    0.89 &    0.36 &     0.93 \\
600/750  &       0.37 &       0.09 &    0.48 &    0.15 &      0.6 &        0.4 &       0.16 &    0.56 &    0.22 &     0.72 &        0.4 &       0.16 &    0.56 &    0.22 &     0.72 \\
600/850  &       0.29 &       0.11 &     0.5 &    0.24 &     0.56 &       0.35 &       0.17 &    0.55 &    0.27 &      0.7 &       0.35 &       0.17 &    0.55 &    0.27 &      0.7 \\
600/1000 &       0.31 &       0.17 &    0.31 &    0.19 &     0.48 &       0.35 &       0.22 &    0.38 &    0.23 &     0.53 &       0.35 &       0.22 &    0.38 &    0.23 &     0.53 \\
600/1200 &        0.2 &       0.12 &    0.32 &    0.13 &     0.41 &       0.25 &       0.18 &    0.32 &    0.18 &     0.45 &       0.25 &       0.18 &    0.32 &    0.18 &     0.45 \\
625/750  &       0.35 &       0.06 &    0.55 &    0.14 &     0.62 &       0.39 &       0.14 &     0.6 &    0.18 &      0.7 &       0.39 &       0.14 &     0.6 &    0.18 &      0.7 \\
650/800  &       0.29 &       0.07 &    0.36 &    0.16 &     0.46 &       0.34 &       0.12 &    0.42 &    0.19 &     0.61 &       0.34 &       0.12 &    0.42 &    0.19 &     0.61 \\
650/850  &       0.28 &       0.06 &     0.5 &    0.38 &     0.56 &       0.31 &       0.11 &    0.51 &     0.4 &     0.66 &       0.31 &       0.11 &    0.51 &     0.4 &     0.66 \\
700/705  &       0.24 &       0.51 &    0.61 &     0.5 &     0.74 &       0.32 &       0.56 &     0.7 &    0.53 &     0.85 &       0.32 &       0.56 &     0.7 &    0.53 &     0.85 \\
700/720  &       0.31 &       0.27 &    0.57 &    0.62 &     0.67 &       0.47 &       0.41 &    0.57 &    0.68 &     0.77 &       0.47 &       0.41 &    0.57 &    0.68 &     0.77 \\
800/805  &        0.1 &       0.19 &    0.22 &    0.24 &     0.32 &       0.14 &        0.3 &    0.33 &    0.48 &     0.57 &       0.14 &        0.3 &    0.33 &    0.48 &     0.57 \\
800/850  &       0.08 &       0.11 &    0.34 &    0.18 &     0.37 &       0.13 &       0.13 &    0.35 &     0.2 &     0.47 &       0.13 &       0.13 &    0.35 &     0.2 &     0.47 \\
800/1200 &       0.09 &       0.04 &     0.1 &    0.05 &     0.14 &       0.12 &       0.06 &    0.13 &    0.07 &     0.19 &       0.12 &       0.06 &    0.13 &    0.07 &     0.19 \\
\end{tabular}
\caption{Subset of expected \glspl*{CL} for squark-pair production ($\tilde{q}\tilde{q}$) only, and after combining with neutralino-squark production ($\tilde{q}\tilde{q}-\tilde{\chi}\tilde{q}$) and neutralino pair-production ($\tilde{q}\tilde{q}-\tilde{\chi}\tilde{q}-\tilde{\chi}\tilde{\chi}$). We consider the ATLAS-CONF-2019-040 (AC), ATLAS-EXOT-2018-06 (AE), CMS-SUS-19-006 (CS) and CMS-EXO-20-004 (CE) analyses, as well as their combination (\textit{cmb}.).\label{sec:cls_exp}}
\end{table}

\begin{table}
    \centering
    \tiny{}
\begin{tabular}{c|lllll|lllll|lllll}
{$m_{\neutralino{1}}$/$m_{\squark{q}}$} & \multicolumn{5}{c}{$\tilde{c}\tilde{q}$} & \multicolumn{5}{c}{$\tilde{q}\tilde{q}-\tilde{\chi}\tilde{q}$} & \multicolumn{5}{c}{$\tilde{q}\tilde{q}-\tilde{\chi}\tilde{q}-\tilde{\chi}\tilde{\chi}$} \\[.2cm]
{GeV} & AC & AE & CS & CE & \textit{cmb}. & AC & AE & CS & CE & \textit{cmb}.& AC & AE & CS & CE & \textit{cmb}.\\
\midrule
200/900  &       0.96 &       0.63 &     1.0 &    0.61 &      1.0 &       0.99 &       0.83 &     1.0 &    0.76 &      1.0 &       0.99 &       0.83 &     1.0 &    0.76 &      1.0 \\
200/950  &       0.94 &       0.64 &    0.97 &    0.57 &      1.0 &       0.98 &        0.8 &     1.0 &    0.69 &      1.0 &       0.98 &        0.8 &    0.99 &    0.69 &      1.0 \\
200/970  &       0.92 &       0.64 &    0.96 &    0.65 &     0.99 &       0.98 &        0.8 &    0.99 &    0.91 &      1.0 &       0.98 &       0.81 &    0.99 &    0.91 &      1.0 \\
200/1000 &       0.91 &       0.68 &    0.94 &    0.69 &     0.99 &       0.96 &       0.83 &    0.98 &    0.76 &      1.0 &       0.96 &       0.84 &    0.99 &    0.76 &      1.0 \\
200/1050 &       0.85 &       0.63 &    0.89 &     0.7 &     0.97 &       0.92 &        0.8 &    0.97 &    0.81 &      1.0 &       0.92 &       0.81 &    0.97 &    0.81 &      1.0 \\
200/1200 &       0.74 &       0.42 &     0.7 &    0.48 &     0.88 &       0.81 &       0.59 &    0.88 &    0.69 &     0.97 &       0.81 &        0.6 &    0.88 &    0.69 &     0.97 \\
200/1300 &       0.69 &       0.43 &     1.0 &    0.38 &     0.82 &        0.8 &        0.6 &     1.0 &     0.7 &     0.95 &        0.8 &       0.61 &     1.0 &     0.7 &     0.95 \\
250/750  &       0.94 &       0.54 &     1.0 &    0.84 &      1.0 &       0.98 &       0.81 &     1.0 &    0.94 &      1.0 &       0.98 &       0.81 &     1.0 &    0.94 &      1.0 \\
250/850  &       0.97 &       0.54 &    0.99 &    0.74 &      1.0 &       0.99 &       0.72 &     1.0 &    0.86 &      1.0 &       0.99 &       0.72 &     1.0 &    0.86 &      1.0 \\
250/940  &       0.93 &       0.55 &    0.99 &    0.59 &      1.0 &       0.97 &       0.76 &    0.99 &    0.77 &      1.0 &       0.97 &       0.76 &    0.99 &    0.77 &      1.0 \\
250/950  &       0.91 &       0.59 &    0.96 &    0.65 &     0.99 &       0.96 &       0.75 &    0.99 &    0.73 &      1.0 &       0.96 &       0.75 &    0.99 &    0.73 &      1.0 \\
250/1000 &        0.9 &       0.57 &    0.94 &    0.67 &     0.99 &       0.95 &       0.77 &     1.0 &    0.79 &      1.0 &       0.95 &       0.77 &     1.0 &     0.8 &      1.0 \\
250/1050 &       0.85 &       0.63 &    0.89 &     0.7 &     0.99 &       0.91 &       0.79 &    0.96 &     0.8 &      1.0 &       0.91 &       0.79 &    0.96 &     0.8 &      1.0 \\
250/1080 &       0.79 &       0.58 &    0.85 &    0.61 &     0.96 &       0.87 &       0.74 &    0.95 &    0.74 &     0.99 &       0.87 &       0.74 &    0.95 &    0.74 &     0.99 \\
250/1100 &       0.76 &       0.54 &    0.83 &    0.62 &     0.95 &       0.85 &        0.7 &    0.94 &    0.72 &     0.98 &       0.85 &        0.7 &    0.94 &    0.72 &     0.98 \\
300/750  &       0.88 &       0.48 &     1.0 &    0.83 &      1.0 &       0.94 &       0.71 &     1.0 &     0.9 &      1.0 &       0.94 &       0.72 &     1.0 &     0.9 &      1.0 \\
300/850  &       0.91 &       0.44 &    0.99 &     0.7 &      1.0 &       0.95 &       0.64 &     1.0 &    0.82 &      1.0 &       0.96 &       0.64 &     1.0 &    0.82 &      1.0 \\
300/900  &       0.93 &       0.51 &    0.98 &    0.57 &      1.0 &       0.97 &       0.67 &     1.0 &    0.71 &      1.0 &       0.97 &       0.67 &     1.0 &    0.72 &      1.0 \\
300/1000 &       0.85 &       0.52 &    0.92 &    0.66 &     0.98 &       0.91 &       0.69 &    0.97 &    0.74 &     0.99 &       0.91 &       0.69 &    0.98 &    0.74 &      1.0 \\
300/1200 &       0.64 &       0.38 &    0.69 &     0.5 &     0.72 &       0.72 &       0.54 &    0.86 &    0.61 &     0.95 &       0.72 &       0.54 &    0.86 &    0.61 &     0.95 \\
350/650  &       0.88 &       0.32 &     1.0 &    0.86 &      1.0 &       0.95 &       0.55 &     1.0 &    0.92 &      1.0 &       0.95 &       0.55 &     1.0 &    0.92 &      1.0 \\
350/700  &       0.76 &       0.45 &    0.99 &    0.83 &      1.0 &       0.87 &       0.62 &     1.0 &    0.89 &      1.0 &       0.87 &       0.62 &     1.0 &    0.89 &      1.0 \\
350/800  &       0.82 &       0.38 &    0.98 &    0.75 &      1.0 &       0.93 &       0.59 &     1.0 &    0.85 &      1.0 &       0.93 &       0.59 &     1.0 &    0.85 &      1.0 \\
350/850  &       0.84 &       0.33 &    0.98 &    0.64 &      1.0 &       0.91 &       0.53 &    0.99 &    0.74 &      1.0 &       0.91 &       0.53 &    0.99 &    0.74 &      1.0 \\
350/900  &       0.86 &       0.42 &    0.96 &    0.55 &     0.98 &       0.92 &       0.57 &    0.99 &    0.68 &      1.0 &       0.92 &       0.58 &    0.99 &    0.68 &      1.0 \\
350/1000 &       0.83 &       0.45 &    0.91 &    0.55 &     0.98 &       0.88 &       0.59 &    0.96 &    0.64 &     0.99 &       0.88 &        0.6 &    0.96 &    0.64 &     0.99 \\
350/1100 &       0.72 &       0.45 &    0.81 &    0.59 &     0.92 &       0.79 &       0.57 &     0.9 &    0.64 &     0.97 &        0.8 &       0.58 &     0.9 &    0.64 &     0.97 \\
400/600  &       0.73 &       0.32 &     1.0 &    0.88 &      1.0 &       0.92 &       0.52 &     1.0 &    0.91 &      1.0 &       0.92 &       0.52 &     1.0 &    0.91 &      1.0 \\
400/750  &       0.84 &       0.24 &    0.97 &    0.62 &     0.99 &       0.88 &        0.4 &    0.99 &    0.72 &      1.0 &       0.88 &        0.4 &    0.99 &    0.72 &      1.0 \\
400/800  &       0.64 &       0.31 &    0.94 &    0.67 &     0.99 &       0.77 &       0.48 &    0.98 &    0.74 &     0.99 &       0.77 &       0.49 &    0.98 &    0.74 &     0.99 \\
400/850  &       0.76 &        0.3 &    0.95 &    0.52 &     0.98 &       0.85 &       0.48 &    0.98 &    0.63 &      1.0 &       0.85 &       0.49 &    0.98 &    0.63 &      1.0 \\
400/870  &       0.74 &       0.28 &    0.95 &    0.58 &     0.98 &       0.83 &       0.44 &    0.98 &    0.72 &      1.0 &       0.83 &       0.45 &    0.98 &    0.72 &      1.0 \\
400/920  &       0.78 &        0.3 &    0.94 &    0.53 &     0.98 &       0.85 &       0.44 &    0.97 &    0.62 &     0.99 &       0.85 &       0.44 &    0.97 &    0.63 &     0.99 \\
400/950  &       0.79 &       0.35 &    0.92 &    0.45 &     0.98 &       0.87 &       0.54 &    0.97 &    0.58 &     0.99 &       0.87 &       0.54 &    0.97 &    0.59 &     0.99 \\
400/1000 &       0.76 &       0.33 &    0.87 &    0.86 &     0.97 &       0.81 &       0.48 &    0.94 &    0.89 &     0.99 &       0.81 &       0.49 &    0.94 &    0.89 &     0.99 \\
400/1200 &       0.56 &       0.36 &    0.69 &    0.41 &     0.81 &       0.67 &       0.49 &    0.83 &    0.52 &     0.93 &       0.67 &       0.49 &    0.83 &    0.52 &     0.93 \\
450/550  &       0.38 &       0.38 &     1.0 &    0.54 &      1.0 &       0.38 &        0.6 &     1.0 &    0.78 &      1.0 &       0.38 &        0.6 &     1.0 &    0.78 &      1.0 \\
450/600  &       0.38 &       0.23 &     1.0 &    0.56 &      1.0 &       0.38 &       0.35 &     1.0 &    0.61 &      1.0 &       0.38 &       0.36 &     1.0 &    0.61 &      1.0 \\
450/650  &       0.59 &       0.22 &     1.0 &    0.48 &      1.0 &        0.7 &       0.37 &     1.0 &    0.56 &      1.0 &        0.7 &       0.37 &     1.0 &    0.57 &      1.0 \\
450/820  &       0.38 &       0.18 &    0.86 &    0.48 &     0.93 &       0.38 &       0.38 &    0.93 &    0.56 &     0.99 &       0.38 &       0.38 &    0.93 &    0.56 &      1.0 \\
450/850  &       0.38 &       0.36 &     0.9 &    0.52 &     0.96 &       0.38 &       0.47 &    0.95 &    0.57 &     0.98 &       0.38 &       0.47 &    0.95 &    0.57 &     0.98 \\
450/950  &       0.38 &       0.26 &     0.9 &    0.46 &     0.95 &       0.38 &       0.39 &    0.94 &    0.57 &     0.99 &       0.38 &       0.39 &    0.94 &    0.57 &     0.99 \\
500/520  &       0.83 &       0.44 &     1.0 &    0.92 &      1.0 &       0.87 &       0.59 &     1.0 &    0.94 &      1.0 &       0.87 &       0.59 &     1.0 &    0.94 &      1.0 \\
500/600  &       0.38 &       0.18 &     1.0 &    0.49 &      1.0 &       0.38 &       0.33 &     1.0 &    0.64 &      1.0 &       0.38 &       0.34 &     1.0 &    0.64 &      1.0 \\
500/650  &       0.38 &       0.31 &    0.82 &    0.43 &     0.96 &       0.38 &       0.43 &    0.91 &     0.5 &     0.97 &       0.38 &       0.43 &    0.91 &     0.5 &     0.97 \\
500/700  &       0.38 &       0.14 &    0.88 &    0.41 &     0.94 &       0.38 &       0.34 &    0.89 &    0.47 &     0.95 &       0.38 &       0.35 &    0.89 &    0.47 &     0.95 \\
500/750  &       0.38 &       0.18 &     0.8 &    0.51 &     0.88 &       0.38 &       0.33 &    0.87 &    0.56 &      0.9 &       0.38 &       0.33 &    0.87 &    0.56 &      0.9 \\
500/800  &       0.38 &       0.19 &    0.74 &    0.58 &     0.74 &       0.38 &       0.34 &    0.85 &    0.63 &     0.93 &       0.38 &       0.34 &    0.85 &    0.63 &     0.94 \\
500/850  &        0.5 &       0.15 &     0.8 &    0.48 &     0.93 &       0.58 &        0.3 &    0.88 &    0.54 &     0.92 &       0.58 &        0.3 &    0.88 &    0.54 &     0.92 \\
500/950  &       0.58 &       0.22 &    0.84 &     0.4 &     0.93 &       0.66 &       0.32 &     0.9 &    0.48 &     0.96 &       0.66 &       0.32 &     0.9 &    0.48 &     0.96 \\
550/700  &        0.4 &       0.17 &    0.69 &    0.34 &     0.77 &       0.44 &       0.27 &    0.81 &    0.41 &     0.87 &       0.44 &       0.28 &    0.81 &    0.41 &     0.87 \\
550/750  &       0.39 &       0.17 &    0.69 &    0.43 &     0.79 &       0.47 &        0.3 &    0.79 &    0.46 &     0.89 &       0.47 &        0.3 &    0.79 &    0.46 &     0.89 \\
550/800  &       0.54 &       0.12 &    0.79 &    0.36 &     0.87 &       0.62 &       0.38 &    0.82 &    0.42 &     0.93 &       0.62 &       0.39 &    0.82 &    0.42 &     0.93 \\
550/850  &       0.47 &       0.13 &    0.85 &    0.39 &     0.88 &       0.52 &       0.24 &    0.85 &    0.45 &     0.85 &       0.52 &       0.24 &    0.85 &    0.45 &     0.85 \\
550/900  &        0.4 &       0.12 &    0.73 &    0.32 &     0.83 &       0.49 &       0.21 &    0.82 &    0.39 &     0.84 &       0.49 &       0.21 &    0.82 &    0.39 &     0.84 \\
600/650  &       0.54 &       0.34 &    0.97 &    0.37 &     0.98 &       0.58 &       0.42 &    0.98 &    0.42 &     0.98 &       0.58 &       0.42 &    0.98 &    0.42 &     0.98 \\
600/750  &       0.35 &       0.16 &    0.64 &    0.31 &     0.63 &       0.38 &       0.29 &    0.76 &    0.38 &     0.83 &       0.38 &        0.3 &    0.76 &    0.38 &     0.83 \\
600/850  &       0.32 &       0.21 &    0.62 &    0.45 &     0.68 &       0.41 &       0.29 &    0.73 &    0.48 &     0.84 &       0.41 &       0.29 &    0.73 &    0.48 &     0.84 \\
600/1000 &        0.4 &       0.12 &    0.68 &    0.27 &     0.73 &       0.46 &       0.18 &    0.76 &    0.33 &     0.78 &       0.46 &       0.19 &    0.76 &    0.33 &     0.78 \\
600/1200 &       0.44 &       0.21 &    0.61 &    0.26 &     0.72 &       0.49 &       0.28 &    0.71 &    0.33 &      0.8 &       0.49 &       0.28 &    0.72 &    0.33 &      0.8 \\
625/750  &       0.33 &       0.17 &    0.62 &    0.22 &     0.75 &       0.37 &       0.28 &    0.73 &    0.27 &     0.79 &       0.37 &       0.29 &    0.73 &    0.27 &     0.79 \\
650/800  &       0.27 &        0.1 &    0.57 &    0.22 &     0.63 &       0.32 &        0.2 &    0.67 &    0.26 &     0.73 &       0.32 &        0.2 &    0.67 &    0.26 &     0.73 \\
650/850  &       0.28 &        0.1 &    0.75 &    0.37 &     0.84 &       0.33 &       0.17 &    0.75 &     0.4 &     0.84 &       0.33 &       0.17 &    0.75 &     0.4 &     0.84 \\
700/705  &       0.48 &       0.37 &    0.84 &    0.42 &     0.89 &        0.5 &       0.52 &    0.88 &    0.45 &     0.91 &        0.5 &       0.52 &    0.88 &    0.45 &     0.91 \\
700/720  &       0.39 &       0.26 &    0.72 &    0.64 &     0.77 &       0.55 &        0.4 &    0.82 &     0.7 &     0.88 &       0.55 &        0.4 &    0.82 &     0.7 &     0.88 \\
800/805  &       0.22 &       0.27 &    0.62 &    0.23 &     0.58 &       0.28 &       0.48 &    0.68 &    0.35 &      0.8 &       0.28 &       0.48 &    0.68 &    0.35 &      0.8 \\
800/850  &       0.25 &       0.19 &    0.58 &    0.33 &     0.61 &       0.27 &       0.26 &    0.64 &    0.34 &     0.68 &       0.27 &       0.26 &    0.64 &    0.34 &     0.68 \\
800/1200 &       0.23 &       0.08 &    0.42 &    0.13 &     0.48 &       0.27 &       0.13 &     0.5 &    0.16 &     0.56 &       0.27 &       0.13 &     0.5 &    0.16 &     0.56 \\
\end{tabular}
\caption{Same as in \cref{sec:cls_exp} but for observed exclusions.}
\end{table}

\clearpage
\bibliography{References}

\providecommand{\href}[2]{#2}\begingroup\raggedright\begin{thebibliography}{10}

\bibitem{PhysRev.159.1251}
S.~Coleman and J.~Mandula, {\it All possible symmetries of the $s$ matrix},
  {\em Phys. Rev.} {\bf 159} (Jul, 1967) 1251--1256.

\bibitem{Haag:1974qh}
R.~Haag, J.~T. Lopuszanski, and M.~Sohnius, {\it {All Possible Generators of
  Supersymmetries of the s Matrix}},  {\em Nucl. Phys. B} {\bf 88} (1975) 257.

\bibitem{Wess:1974tw}
J.~Wess and B.~Zumino, {\it {Supergauge Transformations in Four-Dimensions}},
  {\em Nucl. Phys. B} {\bf 70} (1974) 39--50.

\bibitem{Farrar:1978xj}
G.~R. Farrar and P.~Fayet, {\it {Phenomenology of the Production, Decay, and
  Detection of New Hadronic States Associated with Supersymmetry}},  {\em Phys.
  Lett. B} {\bf 76} (1978) 575--579.

\bibitem{Jungman:1995df}
G.~Jungman, M.~Kamionkowski, and K.~Griest, {\it {Supersymmetric dark matter}},
   {\em Phys. Rept.} {\bf 267} (1996) 195--373,
  [\href{http://arxiv.org/abs/hep-ph/9506380}{{\tt hep-ph/9506380}}].

\bibitem{ATLAS:2011xeq}
{\bf ATLAS} Collaboration, G.~Aad et~al., {\it {Search for supersymmetry using
  final states with one lepton, jets, and missing transverse momentum with the
  ATLAS detector in $\sqrt{s}=7$ TeV $pp$}},  {\em Phys. Rev. Lett.} {\bf 106}
  (2011) 131802, [\href{http://arxiv.org/abs/1102.2357}{{\tt
  arXiv:1102.2357}}].

\bibitem{CMS:2011xek}
{\bf CMS} Collaboration, V.~Khachatryan et~al., {\it {Search for Supersymmetry
  in pp Collisions at 7 TeV in Events with Jets and Missing Transverse
  Energy}},  {\em Phys. Lett. B} {\bf 698} (2011) 196--218,
  [\href{http://arxiv.org/abs/1101.1628}{{\tt arXiv:1101.1628}}].

\bibitem{ATLAS:2012uah}
{\bf ATLAS} Collaboration, G.~Aad et~al., {\it {Search for direct slepton and
  gaugino production in final states with two leptons and missing transverse
  momentum with the ATLAS detector in $pp$ collisions at $\sqrt{s}=7$ TeV}},
  {\em Phys. Lett. B} {\bf 718} (2013) 879--901,
  [\href{http://arxiv.org/abs/1208.2884}{{\tt arXiv:1208.2884}}].

\bibitem{ATLAS:2014ikz}
{\bf ATLAS} Collaboration, G.~Aad et~al., {\it {Search for direct production of
  charginos and neutralinos in events with three leptons and missing transverse
  momentum in $\sqrt{s} =$ 8TeV $pp$ collisions with the ATLAS detector}},
  {\em JHEP} {\bf 04} (2014) 169, [\href{http://arxiv.org/abs/1402.7029}{{\tt
  arXiv:1402.7029}}].

\bibitem{CMS:2012hnc}
{\bf CMS} Collaboration, S.~Chatrchyan et~al., {\it {Search for electroweak
  production of charginos and neutralinos using leptonic final states in $pp$
  collisions at $\sqrt{s}=7$ TeV}},  {\em JHEP} {\bf 11} (2012) 147,
  [\href{http://arxiv.org/abs/1209.6620}{{\tt arXiv:1209.6620}}].

\bibitem{CMS:2013bda}
{\bf CMS} Collaboration, {\it {Search for direct EWK production of SUSY
  particles in multilepton modes with 8TeV data}},  CMS-PAS-SUS-12-022.

\bibitem{ATLAS:2021fbt}
{\bf ATLAS} Collaboration, G.~Aad et~al., {\it {Search for R-parity-violating
  supersymmetry in a final state containing leptons and many jets with the
  ATLAS experiment using $\sqrt{s} = 13 { TeV}$ proton\textendash{}proton
  collision data}},  {\em Eur. Phys. J. C} {\bf 81} (2021), no.~11 1023,
  [\href{http://arxiv.org/abs/2106.09609}{{\tt arXiv:2106.09609}}].

\bibitem{ATLAS:2023afl}
{\bf ATLAS} Collaboration, G.~Aad et~al., {\it {Search for pair production of
  squarks or gluinos decaying via sleptons or weak bosons in final states with
  two same-sign or three leptons with the ATLAS detector}},  {\em JHEP} {\bf
  02} (2024) 107, [\href{http://arxiv.org/abs/2307.01094}{{\tt
  arXiv:2307.01094}}].

\bibitem{ATLAS:2024fyl}
{\bf ATLAS} Collaboration, G.~Aad et~al., {\it {ATLAS Run 2 searches for
  electroweak production of supersymmetric particles interpreted within the
  pMSSM}},  \href{http://arxiv.org/abs/2402.01392}{{\tt arXiv:2402.01392}}.

\bibitem{CMS:2023xlp}
{\bf CMS} Collaboration, A.~Hayrapetyan et~al., {\it {Search for new physics in
  multijet events with at least one photon and large missing transverse
  momentum in proton-proton collisions at 13 TeV}},  {\em JHEP} {\bf 10} (2023)
  046, [\href{http://arxiv.org/abs/2307.16216}{{\tt arXiv:2307.16216}}].

\bibitem{CMS:2023yzg}
{\bf CMS} Collaboration, A.~Tumasyan et~al., {\it {Search for top squark pair
  production in a final state with at least one hadronically decaying tau
  lepton in proton-proton collisions at $ \sqrt{s} $ = 13 TeV}},  {\em JHEP}
  {\bf 07} (2023) 110, [\href{http://arxiv.org/abs/2304.07174}{{\tt
  arXiv:2304.07174}}].

\bibitem{Lara:2022new}
I.~n. Lara, T.~Buanes, R.~Mase\l{}ek, M.~M. Nojiri, K.~Rolbiecki, and
  K.~Sakurai, {\it {Monojet signatures from gluino and squark decays}},  {\em
  JHEP} {\bf 10} (2022) 150, [\href{http://arxiv.org/abs/2208.01651}{{\tt
  arXiv:2208.01651}}].

\bibitem{ATL-PHYS-PUB-2019-029}
{\bf ATLAS} Collaboration, {ATLAS Collaboration}, {\it {Reproducing searches
  for new physics with the ATLAS experiment through publication of full
  statistical likelihoods}},  Tech. Rep.
  \href{http://cds.cern.ch/record/2684863}{ATL-PHYS-PUB-2019-029}, 2019.

\bibitem{CMS-NOTE-2017-001}
{\bf CMS} Collaboration, {CMS Collaboration}, {\it {Simplified likelihood for
  the re-interpretation of public CMS results}},  Tech. Rep.
  \href{https://cds.cern.ch/record/2242860}{CMS-NOTE-2017-001}, CERN, Geneva,
  Jan, 2017.

\bibitem{Araz:2022vtr}
J.~Y. Araz, A.~Buckley, B.~Fuks, H.~Reyes-Gonzalez, W.~Waltenberger, S.~L.
  Williamson, and J.~Yellen, {\it {Strength in numbers: Optimal and scalable
  combination of LHC new-physics searches}},  {\em SciPost Phys.} {\bf 14}
  (2023), no.~4 077, [\href{http://arxiv.org/abs/2209.00025}{{\tt
  arXiv:2209.00025}}].

\bibitem{Altakach:2023tsd}
M.~M. Altakach, S.~Kraml, A.~Lessa, S.~Narasimha, T.~Pascal, T.~Reymermier, and
  W.~Waltenberger, {\it {Global LHC constraints on electroweak-inos with
  SModelS v2.3}},  \href{http://arxiv.org/abs/2312.16635}{{\tt
  arXiv:2312.16635}}.

\bibitem{Alguero:2022gwm}
G.~Alguero, J.~Y. Araz, B.~Fuks, and S.~Kraml, {\it {Signal region combination
  with full and simplified likelihoods in MadAnalysis 5}},  {\em SciPost Phys.}
  {\bf 14} (2023), no.~1 009, [\href{http://arxiv.org/abs/2206.14870}{{\tt
  arXiv:2206.14870}}].

\bibitem{Alwall:2014hca}
J.~Alwall, R.~Frederix, S.~Frixione, V.~Hirschi, F.~Maltoni, O.~Mattelaer,
  H.~S. Shao, T.~Stelzer, P.~Torrielli, and M.~Zaro, {\it {The automated
  computation of tree-level and next-to-leading order differential cross
  sections, and their matching to parton shower simulations}},  {\em JHEP} {\bf
  07} (2014) 079, [\href{http://arxiv.org/abs/1405.0301}{{\tt
  arXiv:1405.0301}}].

\bibitem{Darme:2023jdn}
L.~Darm\'e et~al., {\it {UFO 2.0: the \textquoteleft{}Universal Feynman
  Output\textquoteright{} format}},  {\em Eur. Phys. J. C} {\bf 83} (2023),
  no.~7 631, [\href{http://arxiv.org/abs/2304.09883}{{\tt arXiv:2304.09883}}].

\bibitem{Duhr:2011se}
C.~Duhr and B.~Fuks, {\it {A superspace module for the FeynRules package}},
  {\em Comput. Phys. Commun.} {\bf 182} (2011) 2404--2426,
  [\href{http://arxiv.org/abs/1102.4191}{{\tt arXiv:1102.4191}}].

\bibitem{Christensen:2009jx}
N.~D. Christensen, P.~de~Aquino, C.~Degrande, C.~Duhr, B.~Fuks, M.~Herquet,
  F.~Maltoni, and S.~Schumann, {\it {A Comprehensive approach to new physics
  simulations}},  {\em Eur. Phys. J. C} {\bf 71} (2011) 1541,
  [\href{http://arxiv.org/abs/0906.2474}{{\tt arXiv:0906.2474}}].

\bibitem{Alloul:2013bka}
A.~Alloul, N.~D. Christensen, C.~Degrande, C.~Duhr, and B.~Fuks, {\it
  {FeynRules 2.0 - A complete toolbox for tree-level phenomenology}},  {\em
  Comput. Phys. Commun.} {\bf 185} (2014) 2250--2300,
  [\href{http://arxiv.org/abs/1310.1921}{{\tt arXiv:1310.1921}}].

\bibitem{Bailey:2020ooq}
S.~Bailey, T.~Cridge, L.~A. Harland-Lang, A.~D. Martin, and R.~S. Thorne, {\it
  {Parton distributions from LHC, HERA, Tevatron and fixed target data: MSHT20
  PDFs}},  {\em Eur. Phys. J. C} {\bf 81} (2021), no.~4 341,
  [\href{http://arxiv.org/abs/2012.04684}{{\tt arXiv:2012.04684}}].

\bibitem{Buckley:2014ana}
A.~Buckley, J.~Ferrando, S.~Lloyd, K.~Nordstr\"om, B.~Page, M.~R\"ufenacht,
  M.~Sch\"onherr, and G.~Watt, {\it {LHAPDF6: parton density access in the LHC
  precision era}},  {\em Eur. Phys. J. C} {\bf 75} (2015) 132,
  [\href{http://arxiv.org/abs/1412.7420}{{\tt arXiv:1412.7420}}].

\bibitem{Fiaschi:2023tkq}
J.~Fiaschi, B.~Fuks, M.~Klasen, and A.~Neuwirth, {\it {Electroweak superpartner
  production at 13.6 Tev with Resummino}},  {\em Eur. Phys. J. C} {\bf 83}
  (2023), no.~8 707, [\href{http://arxiv.org/abs/2304.11915}{{\tt
  arXiv:2304.11915}}].

\bibitem{Beenakker:2016lwe}
W.~Beenakker, C.~Borschensky, M.~Kr\"amer, A.~Kulesza, and E.~Laenen, {\it
  {NNLL-fast: predictions for coloured supersymmetric particle production at
  the LHC with threshold and Coulomb resummation}},  {\em JHEP} {\bf 12} (2016)
  133, [\href{http://arxiv.org/abs/1607.07741}{{\tt arXiv:1607.07741}}].

\bibitem{HEPi}
A.~P. Neuwirth, {\it Apn-pucky/hepi: Zenodo release},  Oct., 2023.

\bibitem{Sterman:1986aj}
G.~F. Sterman, {\it {Summation of Large Corrections to Short Distance Hadronic
  Cross-Sections}},  {\em Nucl. Phys. B} {\bf 281} (1987) 310--364.

\bibitem{Catani:1989ne}
S.~Catani and L.~Trentadue, {\it {Resummation of the QCD Perturbative Series
  for Hard Processes}},  {\em Nucl. Phys. B} {\bf 327} (1989) 323--352.

\bibitem{Vogt:2000ci}
A.~Vogt, {\it {Next-to-next-to-leading logarithmic threshold resummation for
  deep inelastic scattering and the Drell-Yan process}},  {\em Phys. Lett. B}
  {\bf 497} (2001) 228--234, [\href{http://arxiv.org/abs/hep-ph/0010146}{{\tt
  hep-ph/0010146}}].

\bibitem{Beenakker:1999xh}
W.~Beenakker, M.~Klasen, M.~Kramer, T.~Plehn, M.~Spira, and P.~M. Zerwas, {\it
  {The Production of charginos / neutralinos and sleptons at hadron
  colliders}},  {\em Phys. Rev. Lett.} {\bf 83} (1999) 3780--3783,
  [\href{http://arxiv.org/abs/hep-ph/9906298}{{\tt hep-ph/9906298}}]. [Erratum:
  Phys.Rev.Lett. 100, 029901 (2008)].

\bibitem{Fiaschi:2020udf}
J.~Fiaschi and M.~Klasen, {\it {Higgsino and gaugino pair production at the LHC
  with aNNLO+NNLL precision}},  {\em Phys. Rev. D} {\bf 102} (2020), no.~9
  095021, [\href{http://arxiv.org/abs/2006.02294}{{\tt arXiv:2006.02294}}].

\bibitem{Debove:2010kf}
J.~Debove, B.~Fuks, and M.~Klasen, {\it {Threshold resummation for gaugino pair
  production at hadron colliders}},  {\em Nucl. Phys. B} {\bf 842} (2011)
  51--85, [\href{http://arxiv.org/abs/1005.2909}{{\tt arXiv:1005.2909}}].

\bibitem{Fuks:2012qx}
B.~Fuks, M.~Klasen, D.~R. Lamprea, and M.~Rothering, {\it {Gaugino production
  in proton-proton collisions at a center-of-mass energy of 8 TeV}},  {\em
  JHEP} {\bf 10} (2012) 081, [\href{http://arxiv.org/abs/1207.2159}{{\tt
  arXiv:1207.2159}}].

\bibitem{Fiaschi:2018hgm}
J.~Fiaschi and M.~Klasen, {\it {Neutralino-chargino pair production at NLO+NLL
  with resummation-improved parton density functions for LHC Run II}},  {\em
  Phys. Rev. D} {\bf 98} (2018), no.~5 055014,
  [\href{http://arxiv.org/abs/1805.11322}{{\tt arXiv:1805.11322}}].

\bibitem{Baglio:2021zjm}
J.~Baglio, G.~Coniglio, B.~Jager, and M.~Spira, {\it {Next-to-leading-order QCD
  corrections and parton-shower effects for weakino+squark production at the
  LHC}},  {\em JHEP} {\bf 12} (2021) 020,
  [\href{http://arxiv.org/abs/2110.04211}{{\tt arXiv:2110.04211}}].

\bibitem{Fiaschi:2022odp}
J.~Fiaschi, B.~Fuks, M.~Klasen, and A.~Neuwirth, {\it {Soft gluon resummation
  for associated squark-electroweakino production at the LHC}},  {\em JHEP}
  {\bf 06} (2022) 130, [\href{http://arxiv.org/abs/2202.13416}{{\tt
  arXiv:2202.13416}}].

\bibitem{Beenakker:1996ch}
W.~Beenakker, R.~Hopker, M.~Spira, and P.~M. Zerwas, {\it {Squark and gluino
  production at hadron colliders}},  {\em Nucl. Phys. B} {\bf 492} (1997)
  51--103, [\href{http://arxiv.org/abs/hep-ph/9610490}{{\tt hep-ph/9610490}}].

\bibitem{Langenfeld:2009eg}
U.~Langenfeld and S.-O. Moch, {\it {Higher-order soft corrections to squark
  hadro-production}},  {\em Phys. Lett. B} {\bf 675} (2009) 210--221,
  [\href{http://arxiv.org/abs/0901.0802}{{\tt arXiv:0901.0802}}].

\bibitem{Kulesza:2008jb}
A.~Kulesza and L.~Motyka, {\it {Threshold resummation for squark-antisquark and
  gluino-pair production at the LHC}},  {\em Phys. Rev. Lett.} {\bf 102} (2009)
  111802, [\href{http://arxiv.org/abs/0807.2405}{{\tt arXiv:0807.2405}}].

\bibitem{Kulesza:2009kq}
A.~Kulesza and L.~Motyka, {\it {Soft gluon resummation for the production of
  gluino-gluino and squark-antisquark pairs at the LHC}},  {\em Phys. Rev. D}
  {\bf 80} (2009) 095004, [\href{http://arxiv.org/abs/0905.4749}{{\tt
  arXiv:0905.4749}}].

\bibitem{Beenakker:2009ha}
W.~Beenakker, S.~Brensing, M.~Kramer, A.~Kulesza, E.~Laenen, and I.~Niessen,
  {\it {Soft-gluon resummation for squark and gluino hadroproduction}},  {\em
  JHEP} {\bf 12} (2009) 041, [\href{http://arxiv.org/abs/0909.4418}{{\tt
  arXiv:0909.4418}}].

\bibitem{Beenakker:2011sf}
W.~Beenakker, S.~Brensing, M.~Kramer, A.~Kulesza, E.~Laenen, and I.~Niessen,
  {\it {NNLL resummation for squark-antisquark pair production at the LHC}},
  {\em JHEP} {\bf 01} (2012) 076, [\href{http://arxiv.org/abs/1110.2446}{{\tt
  arXiv:1110.2446}}].

\bibitem{Beenakker:2013mva}
W.~Beenakker, T.~Janssen, S.~Lepoeter, M.~Kr\"amer, A.~Kulesza, E.~Laenen,
  I.~Niessen, S.~Thewes, and T.~Van~Daal, {\it {Towards NNLL resummation: hard
  matching coefficients for squark and gluino hadroproduction}},  {\em JHEP}
  {\bf 10} (2013) 120, [\href{http://arxiv.org/abs/1304.6354}{{\tt
  arXiv:1304.6354}}].

\bibitem{Beenakker:2014sma}
W.~Beenakker, C.~Borschensky, M.~Kr\"amer, A.~Kulesza, E.~Laenen, V.~Theeuwes,
  and S.~Thewes, {\it {NNLL resummation for squark and gluino production at the
  LHC}},  {\em JHEP} {\bf 12} (2014) 023,
  [\href{http://arxiv.org/abs/1404.3134}{{\tt arXiv:1404.3134}}].

\bibitem{Beenakker:2015rna}
W.~Beenakker, C.~Borschensky, M.~Kr\"amer, A.~Kulesza, E.~Laenen, S.~Marzani,
  and J.~Rojo, {\it {NLO+NLL squark and gluino production cross-sections with
  threshold-improved parton distributions}},  {\em Eur. Phys. J. C} {\bf 76}
  (2016), no.~2 53, [\href{http://arxiv.org/abs/1510.00375}{{\tt
  arXiv:1510.00375}}].

\bibitem{Beenakker:1996ed}
W.~Beenakker, R.~Hopker, and M.~Spira, {\it {PROSPINO: A Program for the
  production of supersymmetric particles in next-to-leading order QCD}},
  \href{http://arxiv.org/abs/hep-ph/9611232}{{\tt hep-ph/9611232}}.

\bibitem{Gao:2002is}
G.-p. Gao, G.-r. Lu, Z.-h. Xiong, and J.~M. Yang, {\it {Loop effects and
  nondecoupling property of SUSY QCD in g b ---\ensuremath{>} t H-}},  {\em
  Phys. Rev. D} {\bf 66} (2002) 015007,
  [\href{http://arxiv.org/abs/hep-ph/0202016}{{\tt hep-ph/0202016}}].

\bibitem{Bierlich:2022pfr}
C.~Bierlich et~al., {\it {A comprehensive guide to the physics and usage of
  PYTHIA 8.3}},  {\em SciPost Phys. Codeb.} {\bf 2022} (2022) 8,
  [\href{http://arxiv.org/abs/2203.11601}{{\tt arXiv:2203.11601}}].

\bibitem{Catani:2001cc}
S.~Catani, F.~Krauss, R.~Kuhn, and B.~R. Webber, {\it {QCD matrix elements +
  parton showers}},  {\em JHEP} {\bf 11} (2001) 063,
  [\href{http://arxiv.org/abs/hep-ph/0109231}{{\tt hep-ph/0109231}}].

\bibitem{Lonnblad:2001iq}
L.~Lonnblad, {\it {Correcting the color dipole cascade model with fixed order
  matrix elements}},  {\em JHEP} {\bf 05} (2002) 046,
  [\href{http://arxiv.org/abs/hep-ph/0112284}{{\tt hep-ph/0112284}}].

\bibitem{Lonnblad:2011xx}
L.~Lonnblad and S.~Prestel, {\it {Matching Tree-Level Matrix Elements with
  Interleaved Showers}},  {\em JHEP} {\bf 03} (2012) 019,
  [\href{http://arxiv.org/abs/1109.4829}{{\tt arXiv:1109.4829}}].

\bibitem{Conte:2012fm}
E.~Conte, B.~Fuks, and G.~Serret, {\it {MadAnalysis 5, A User-Friendly
  Framework for Collider Phenomenology}},  {\em Comput. Phys. Commun.} {\bf
  184} (2013) 222--256, [\href{http://arxiv.org/abs/1206.1599}{{\tt
  arXiv:1206.1599}}].

\bibitem{Conte:2014zja}
E.~Conte, B.~Dumont, B.~Fuks, and C.~Wymant, {\it {Designing and recasting LHC
  analyses with MadAnalysis 5}},  {\em Eur. Phys. J. C} {\bf 74} (2014), no.~10
  3103, [\href{http://arxiv.org/abs/1405.3982}{{\tt arXiv:1405.3982}}].

\bibitem{Conte:2018vmg}
E.~Conte and B.~Fuks, {\it {Confronting new physics theories to LHC data with
  MADANALYSIS 5}},  {\em Int. J. Mod. Phys. A} {\bf 33} (2018), no.~28 1830027,
  [\href{http://arxiv.org/abs/1808.00480}{{\tt arXiv:1808.00480}}].

\bibitem{deFavereau:2013fsa}
{\bf DELPHES 3} Collaboration, J.~de~Favereau, C.~Delaere, P.~Demin,
  A.~Giammanco, V.~Lema\^\i{}tre, A.~Mertens, and M.~Selvaggi, {\it {DELPHES 3,
  A modular framework for fast simulation of a generic collider experiment}},
  {\em JHEP} {\bf 02} (2014) 057, [\href{http://arxiv.org/abs/1307.6346}{{\tt
  arXiv:1307.6346}}].

\bibitem{Araz:2020lnp}
J.~Y. Araz, B.~Fuks, and G.~Polykratis, {\it {Simplified fast detector
  simulation in MADANALYSIS 5}},  {\em Eur. Phys. J. C} {\bf 81} (2021), no.~4
  329, [\href{http://arxiv.org/abs/2006.09387}{{\tt arXiv:2006.09387}}].

\bibitem{Araz:2023bwx}
J.~Y. Araz, {\it {Spey: smooth inference for reinterpretation studies}},  {\em
  SciPost Phys.} {\bf 16} (2024) 032,
  [\href{http://arxiv.org/abs/2307.06996}{{\tt arXiv:2307.06996}}].

\bibitem{ATLAS:2021kxv}
{\bf ATLAS} Collaboration, G.~Aad et~al., {\it {Search for new phenomena in
  events with an energetic jet and missing transverse momentum in $pp$
  collisions at $\sqrt {s}$ =13 TeV with the ATLAS detector}},  {\em Phys. Rev.
  D} {\bf 103} (2021), no.~11 112006,
  [\href{http://arxiv.org/abs/2102.10874}{{\tt arXiv:2102.10874}}].

\bibitem{ATLAS:2020syg}
{\bf ATLAS} Collaboration, G.~Aad et~al., {\it {Search for squarks and gluinos
  in final states with jets and missing transverse momentum using 139 fb$^{-1}$
  of $\sqrt{s}$ =13 TeV $pp$ collision data with the ATLAS detector}},  {\em
  JHEP} {\bf 02} (2021) 143, [\href{http://arxiv.org/abs/2010.14293}{{\tt
  arXiv:2010.14293}}].

\bibitem{CMS:2019zmd}
{\bf CMS} Collaboration, T.~C. Collaboration et~al., {\it {Search for
  supersymmetry in proton-proton collisions at 13 TeV in final states with jets
  and missing transverse momentum}},  {\em JHEP} {\bf 10} (2019) 244,
  [\href{http://arxiv.org/abs/1908.04722}{{\tt arXiv:1908.04722}}].

\bibitem{CMS:2021far}
{\bf CMS} Collaboration, A.~Tumasyan et~al., {\it {Search for new particles in
  events with energetic jets and large missing transverse momentum in
  proton-proton collisions at $ \sqrt{s} $ = 13 TeV}},  {\em JHEP} {\bf 11}
  (2021) 153, [\href{http://arxiv.org/abs/2107.13021}{{\tt arXiv:2107.13021}}].

\bibitem{DVN/NW3NPG_2021}
F.~Ambrogi, {\it {Implementation of a search for squarks and gluinos in the
  multi-jet + missing energy channel (139 fb-1; 13 TeV; ATLAS-CONF-2019-040)}},
   doi:10.14428/DVN/NW3NPG.

\bibitem{DVN/REPAMM_2023}
D.~Agin, {\it {Implementation of a search for new physics with jets and missing
  transverse energy (139/fb; 13 TeV; ATLAS-EXOT-2018-06)}},
  doi:10.14428/DVN/REPAMM.

\bibitem{DVN/4DEJQM_2020}
M.~Malte, S.~Bein, and J.~Sonneveld, {\it {Re-implementation of a search for
  supersymmetry in the HT/missing HT channel (137 fb-1; CMS-SUSY-19-006)}},
  doi:10.14428/DVN/4DEJQM.

\bibitem{DVN/IRF7ZL_2021}
A.~Albert, {\it {Implementation of a search for new phenomena in events
  featuring energetic jets and missing transverse energy (137 fb-1; 13 TeV;
  CMS-EXO-20-004)}},  doi:10.14428/DVN/IRF7ZL.

\bibitem{Dumont:2014tja}
B.~Dumont, B.~Fuks, S.~Kraml, S.~Bein, G.~Chalons, E.~Conte, S.~Kulkarni,
  D.~Sengupta, and C.~Wymant, {\it {Toward a public analysis database for LHC
  new physics searches using MADANALYSIS 5}},  {\em Eur. Phys. J. C} {\bf 75}
  (2015), no.~2 56, [\href{http://arxiv.org/abs/1407.3278}{{\tt
  arXiv:1407.3278}}].

\bibitem{Araz:2019otb}
J.~Y. Araz, M.~Frank, and B.~Fuks, {\it {Reinterpreting the results of the LHC
  with MadAnalysis 5: uncertainties and higher-luminosity estimates}},  {\em
  Eur. Phys. J. C} {\bf 80} (2020), no.~6 531,
  [\href{http://arxiv.org/abs/1910.11418}{{\tt arXiv:1910.11418}}].

\bibitem{Kim:2020nrg}
J.~Kim, T.~Lee, J.~Kim, and H.~Jang, {\it {Implementation of the
  ATLAS-SUSY-2018-06 analysis in the MadAnalysis 5 framework (electroweakinos
  with Jigsaw variables; 139 fb$^{−1}$)}},  {\em Mod. Phys. Lett. A} {\bf 36}
  (2021), no.~01 2141004.

\bibitem{Mrowietz:2020ztq}
M.~Mrowietz, S.~Bein, and J.~Sonneveld, {\it {Implementation of the
  CMS-SUS-19-006 analysis in the MadAnalysis 5 framework (supersymmetry with
  large hadronic activity and missing transverse energy; 137 fb$^{−1}$)}},
  {\em Mod. Phys. Lett. A} {\bf 36} (2021), no.~01 2141007.

\bibitem{Fuks:2021zbm}
B.~Fuks et~al., {\it {Proceedings of the second MadAnalysis 5 workshop on LHC
  recasting in Korea}},  {\em Mod. Phys. Lett. A} {\bf 36} (2021), no.~01
  2102001, [\href{http://arxiv.org/abs/2101.02245}{{\tt arXiv:2101.02245}}].

\bibitem{Cowan:2010js}
G.~Cowan, K.~Cranmer, E.~Gross, and O.~Vitells, {\it {Asymptotic formulae for
  likelihood-based tests of new physics}},  {\em Eur. Phys. J. C} {\bf 71}
  (2011) 1554, [\href{http://arxiv.org/abs/1007.1727}{{\tt arXiv:1007.1727}}].
  [Erratum: Eur.Phys.J.C 73, 2501 (2013)].

\end{thebibliography}\endgroup

\end{document}